\documentclass{aa}  

\usepackage{graphicx}
\usepackage{subfigure}
\usepackage{txfonts}

\usepackage{multirow}
\usepackage{colortbl}
\usepackage{float}
\usepackage{hyperref}  
\usepackage{stfloats}
\hypersetup{colorlinks=true,linkcolor=[rgb]{1.,0.2,0.2},citecolor=[rgb]{0.1,0.4,1.},filecolor=[rgb]{0.7,0.2,0.2},urlcolor=[rgb]{0.7,0.2,0.2}}

\usepackage{color}
\usepackage{xcolor}
\usepackage{caption}
\usepackage{array}
\usepackage{tabularx} 
\usepackage[toc,page]{appendix}
\definecolor{blue}{rgb}{0., 0., 1}
\definecolor{lightblue}{rgb}{0.1,0.4,1.}

\title{Image simulations of highly magnified clumpy galaxies}
\titlerunning{Image simulations}

\author{
I. ~Mini \inst{\ref{difabo},\ref{inafbo}} \fnmsep\thanks{E-mail: \href{mailto:irene.mini2@unibo.it}{irene.mini2@unibo.it}},  M. ~Meneghetti \inst{\ref{inafbo},\ref{infnbo}}, M. ~Messa \inst{\ref{inafbo}}, L. ~Moscardini\inst{\ref{difabo},\ref{inafbo},\ref{infnbo}},  E. ~Vanzella \inst{\ref{inafbo}},
P. ~Bergamini \inst{\ref{unimi},\ref{inafbo}},
P. ~Rosati \inst{\ref{unife}},
A. ~Zanella \inst{\ref{inafbo}}
}

\institute{
DIFA -- Department of Physics and Astronomy, University of Bologna, Via Gobetti 93/2, 40129 Bologna, Italy \label{difabo} 
\and
INAF -- OAS, Astrophysics and Space Science Observatory Bologna, Via Gobetti 93/3, 40129 Bologna, Italy \label{inafbo} 
\and
INFN -- Sezione di Bologna, Viale Berti Pichat 6/2, 40127 Bologna, Italy
\label{infnbo}
\and
Dipartimento di Fisica, Universit\`a  degli Studi di Milano, via Celoria 16, I-20133 Milano, Italy\label{unimi}
\and 
 Department of Physics and Earth Science, University of Ferrara, via Saragat 1, I--44122, Ferrara, Italy\label{unife}
           }

   \date{Received month day, year; accepted month day, year}

\begin{document}

  \abstract
  {We present ClumPyLen, a Python-based simulator designed to produce realistic mock observations of strongly lensed, high-redshift, clumpy star-forming galaxies. The tool models galaxy components such as disks, bulges, and spiral arms using S\'ersic profiles, and it populates them with stellar clumps whose properties are sampled from physically motivated distributions. ClumPyLen includes the effects of gravitational lensing through user-provided deflection angle maps and simulates realistic observational conditions by accounting for instrumental effects, Point-Spread-Function convolution, sky background, and photon noise. The simulator can support a wide range of filters and instruments; here we focus on HST/ACS, HST/WFC3-IR, and JWST/NIRCam. We demonstrate the capabilities of the code through two examples, including a detailed simulation of the z = 6.145 source Cosmic Archipelago lensed by MACS J0416.1-2403. The simulated images closely match the morphology and limiting magnitudes of real observations. ClumPyLen is designed to explore the detectability of stellar clumps in terms of mass and size, especially in the low-mass regime, and it allows the study of clump blending effects. Thanks to its modular design, the code is highly adaptable to a wide range of scientific goals, including lensing studies, galaxy evolution, and the generation of synthetic datasets for machine learning or forward modeling applications.
  }


   \maketitle

\section{Introduction}
 
The clumpy structure, typical of distant galaxies, was first observed in Hubble Space Telescope (HST) images of unlensed sources in deep fields and was attributed to giant star-forming regions with sizes of $\sim 1$~kpc \citep[e.g.,][]{elmegreen2006,wisnioski2012}.
The increase in intrinsic spatial resolution provided by gravitationally lensed fields allowed these large regions to be broken into smaller sub-components, namely clumps with sizes spanning the range of 10-500 pc \citep[e.g.,][]{johnson2017,Vanzella_2017,cava2018,Vanzella_2019,Mestric2022,Messa2022,Messa2024a}. 
Recently, thanks to the James Webb Space Telescope's (JWST) resolution and sensitivity, high-$z$ stellar clumps have become observable at $\sim100$ pc scales even in unlensed fields \citep[e.g.,][]{Ji2024,kalita2024,kalita2025b,kalita2025a}; when considering the aid of gravitational lensing, JWST can probe clumps down to $< 10$ pc in size and $\rm10^5~M_{\odot}$ in mass, reaching the scales of individual star clusters. 
For instance, the "Cosmic Gems" is a gravitational arc at $z \sim 10.2$ \citep[][]{adamo2024boundstarclustersobserved}, lensed by the galaxy cluster SPT-CL J0615-5746 and initially discovered by HST in the RELICS survey \citep[][]{2018ApJ...864L..22S}. 
Using JWST observations, \citet{adamo2024boundstarclustersobserved} detected five different star clusters located in a region smaller than 70 pc, with intrinsic sizes $\sim 1$ pc and ages younger than 50 Myr.
Other notable examples are, for example, the "Sunrise Arc" at $z \sim 6$ \citep[][]{Vanzella_2023}, where six stellar clusters have been detected (characterized by sizes $\sim 1-20$ pc and ages in the range $1-30$ Myr) and the "Firefly Sparkle" galaxy at $z \sim 8.3$ \citep[][]{mowla2024fireflysparkleearlieststages}, containing ten star clusters (with sizes $<10$ pc and masses $\rm \sim10^5~M_\odot$) whose star formation histories (SFHs) show a recent starburst, probably triggered by the interaction with a nearby galaxy.
The "redder" wavelength range probed by JWST, compared to HST, may also unclose the study of old clumpy systems. Several stellar clumps have been identified  in the "Sparkler" galaxy at $z \sim 1.37$, lensed by the galaxy cluster SMACS 0723; these systems are consistent with being red, old, evolved globular clusters with formation redshift $z_{form} \sim 7-11$ and ages between 1 and 4 Gyr \citep[][]{Mowla_2022,Claeyssens2023,Adamo2023}.

Hence, when the angular resolution increases thanks to more advanced instruments or the magnification granted by gravitational lensing, single stellar clusters are also detectable in the high-redshift Universe.
This is a crucial achievement since it allows us to map the star cluster formation in early galaxies.
Young star clusters (YSCs) are the natural site of formation for massive stars \citep[e.g.][]{Zinnecker2007,PortegiesZwart2010}; their detection at high-$z$ can unveil their role in the reionization process, which took place in the redshift interval between $z=5$ and $z=15$ when the first sources started to form and their photons led to the ionization of the neutral hydrogen permeating the Universe.
Observing compact star-forming regions is also fundamental for our understanding of galaxy growth and evolution since they alter the properties of the interstellar medium of the hosting galaxy thanks to their radiative and mechanical feedback. This interconnection with the host is closely related to the clump formation scenario, which is still debated and may evolve with the host's redshift.
The first suggested way is a formation in situ, due to the fragmentation of the disk of the host galaxy triggered by instabilities \citep[][]{2009MNRAS.397L..64A, 2009ApJ...703..785D, 2010MNRAS.409.1088B}. 
The second scenario sees the clumps originating from minor merger events \citep[][]{2017MNRAS.464..635M, 2019MNRAS.489.2792Z}.

Large statistics are needed to derive physical properties of the stellar clumps and clusters robustly.
In one of the largest attempts, \citet{Claeyssens2025} studied $\sim2000$ individual clumps in 476 galaxies at redshift $z \sim 0.7 - 10$ lensed by the galaxy cluster A2744, using JWST NIRCam observations, and recovering the main clump physical properties, such as their sizes, ages, and masses.
They show an evolution of the clumps' properties with redshift (clumps are, on average, smaller and younger at higher redshifts) and correlations between the properties of the clumps and those of their host galaxy (massive galaxies host more massive and older clumps). This sample of high-$z$ clumps is a factor $\sim10$ larger than the one studied in SMACS0723 by \citet{Claeyssens2023}, which, in turn, doubled the number of compact ($\rm <100$ pc) clumps at $z>1$ known before JWST \citep[see, e.g.][]{Mestric2022, Messa2024a}. 
The current HST and JWST samples of high-$z$ lensed clusters are largely inhomogeneous in terms of sizes and luminosity probed; the large difference in magnification, from one source to the other within a field, makes the estimate of completeness a very complex problem, even when focusing on subsamples within a small redshift range. This complexity limits the possibility of studying trends among systems and across redshifts.

For this work, we aimed to implement a pipeline to simulate lensing effects by intervening mass distributions such as galaxy clusters on high-redshift star-forming galaxies hosting multiple star clusters. The pipeline also allows observations to be simulated with various instruments, such as HST and JWST. 
The pipeline uses functions and methods from the multipurpose lensing library {\sc PyLensLib} \citep{2021LNP...956.....M}, written in Python. 
In this paper, we illustrate the pipeline functionalities (Sects.~\ref{sec:source_lens_planes}~to~\ref{sec:noise}) and discuss some potential usages (Sects.~\ref{sec:virtualobs} to \ref{sec:a2744}).
In particular, this software can be particularly useful for studying the detectability of lensed stellar clusters at any redshift, especially those at the low-mass end of the star cluster mass function \citep[e.g.,][]{Adamo_2020}.
In this study, we consider a stellar clump to be “detected” only if it is visually identifiable in the images. This choice is motivated by the relatively small sample analyzed here, which allows for careful visual inspection. In future studies involving larger datasets or simulations, we plan to complement visual inspection with automatic detection software, such as \texttt{SExtractor} \citep[][]{1996A&AS..117..393B}, following the method described in \citet{Claeyssens2025}, to efficiently and systematically identify clumps.

In this paper, we use the expression star formation clump when referring to a sub-component of a galaxy (typically on physical scales of 10-1000 pc). 
In contrast, the term stellar cluster represents the usual gravitationally bound YSC (with typical sizes <10 pc, e.g. \citealp{Ryon2015,ryon2017,brown2021}). 
Hence, depending on the angular resolution, a clump can be a complex or a single stellar cluster.
We assume a flat cosmology with $\Omega_M = 0.3$, $\Omega_{\Lambda} = 0.7$, and $H_0 = 70$ km s$^{-1}$ Mpc$^{-1}$. 

\section{Source and lens planes}
\label{sec:source_lens_planes}

When simulating lensing effects by a given gravitational lens, we use the {\em thin screen approximation}, that is, we assume that the lens mass distribution can be projected on a plane at redshift $z_l$, perpendicular to an arbitrary optical axis passing through the observer position. Similarly, the source is assumed to lay on a plane perpendicular to the same axis at redshift $z_s$. At the moment, we do not consider the combined effects of multiple mass distributions along the line of sight. This scenario can be easily implemented by means of a multiplane ray-tracing algorithm, which is already implemented in {\sc PyLensLib}. 

We define the angular coordinates $\vec\theta=(\theta_1,\theta_2)$ and $\vec\beta=(\beta_1,\beta_2)$ on the lens and source planes, respectively. Each point on the lens plane can be mapped onto the source plane using the {\em lens equation},
\begin{equation}
\vec\beta = \vec\theta -\vec{\alpha}(\vec\theta)\;, 
\label{eq:lens}
\end{equation}
where $\vec{\alpha}(\vec\theta)$ is the lens deflection angle at position $\vec\theta$.

The lens equation also allows us to relate the surface brightness of the lensed images, $I(\vec\theta)$, to the intrinsic surface brightness of the source, $I_s(\vec\beta)$, using the ray-tracing method.
This technique consists of tracing bundles of light rays through the lens plane at positions ($\theta_1, \theta_2$), where the deflection angles can be read off the deflection angle maps, $\alpha_1(\theta_1, \theta_2)$ and $\alpha_2(\theta_1, \theta_2)$. Then, the arrival position of each ray on the source plane can be derived from the lens equation, and the image surface brightness is finally computed as
\begin{equation}
I(\vec{\theta})=I_s\left[\vec\theta-\vec\alpha(\vec\theta)\right] \;.
\end{equation}

\section{Modeling the sources}
\label{sec:sources}
We aim to generate clumpy galaxies for image simulations, including lensing. 
We model these sources using multiple luminous components, describing the hosts and the stellar clumps. 
Analytical functions describe the brightness profiles of each component.
Specifically, we use the S\'ersic profile \citep{1963BAAA....6...41S}, given by

\begin{equation}
I_s(\beta) = I_{e} \exp{\left\{-b(n) \left[ \left(\frac{\beta}{R_e} \right)^{1/n}-1\right]\right\}},
\label{Sersic_pr}
\end{equation}
where $\beta$ is the angular distance from the source center $(\beta_{s,1},\beta_{s,2})$,
\begin{equation}
\beta=\sqrt{(\beta_1-\beta_{s,1})^2+(\beta_2-\beta_{s,2})^2} \;,
\end{equation}
$I_{e}$ is the surface brightness at the effective radius $R_e$, $n$ is the S\'ersic index and $b$ is a function of the S\'ersic index well approximated by $b(n) = 2n -\frac{1}{3} + \frac{4}{405n}$ \citep{1963BAAA....6...41S}. 
 
On average, bulges and elliptical galaxies have the steepest central profiles, with $2 < n < 10$. Galactic disks have exponential profiles with $ n\sim1$. Bars have flatter central profiles with $n \leq 0.5$, while the peculiar case with $n=0.5$ corresponds to a Gaussian brightness profile, often used to describe stellar clumps \citep[e.g.][]{cava2018}.

\subsection{Host galaxy}
\label{sect:host}
Our simplest description of galaxies hosting stellar clumps assumes that they can be well approximated by dual brightness distributions consisting of a spherical bulge and an elliptical, exponential disk. Each of these components has its own effective surface brightness and radius, $I_{e,{\rm bulge}}$, $R_{e,{\rm bulge}}$ and $I_{e,{\rm disk}}$, $R_{e,{\rm disk}}$. 
For the examples in this work, we use S\'ersic profiles with $n=1$ for the disk and $n=4$ for the central bulge. These values can be adapted to the sources one wishes to simulate.

Thus, for such a galaxy, the total surface brightness distribution on the source plane is 
\begin{eqnarray}
I_s(\vec\beta) & = &  I_{e,{\rm bulge}}\exp{\left\{-b(4)\left[\left(\frac{\beta}{R_{e,{\rm bulge}}}\right)^{1/4}-1\right]\right\}} \nonumber \\ & +& I_{0}\exp{\left(-\frac{\tilde\beta}{R_{h}}\right)} \;,
\label{eq:host_gal}
\end{eqnarray}
where 
$I_0=I_{e,{\rm disk}}\exp{(1.678)}$, and $R_h=R_{e, {\rm disk}}/1.678$. To include ellipticity in the disk surface brightness, we define   
\begin{equation}
\tilde\beta=\sqrt{(\beta_1-\beta_{s,1})^2/q^2+(\beta_2-\beta_{s,2})^2} \;,
\label{eq:betatilde}
\end{equation}
where $q$ is the disk axis ratio. 

We can simulate arbitrary position angles $\varphi$ for the source galaxy by applying a rotation of the reference frame:
\begin{equation}
\begin{gathered}
    \beta_1' = (\beta_1 - \beta_{s,1})  \cos\varphi + (\beta_2 - \beta_{s,2})  \sin\varphi, 
\\
    \beta_2' = -(\beta_1 - \beta_{s,1})  \sin\varphi + (\beta_2 - \beta_{s,2}) \cos\varphi \,,
\end{gathered}
\label{eq:betas}
\end{equation}
where $(\beta_1,\beta_2)$ and $(\beta_1',\beta_2')$ are the original and new  coordinates, respectively.

\begin{figure*}[t]
    \centering    
    \includegraphics[width=\linewidth]{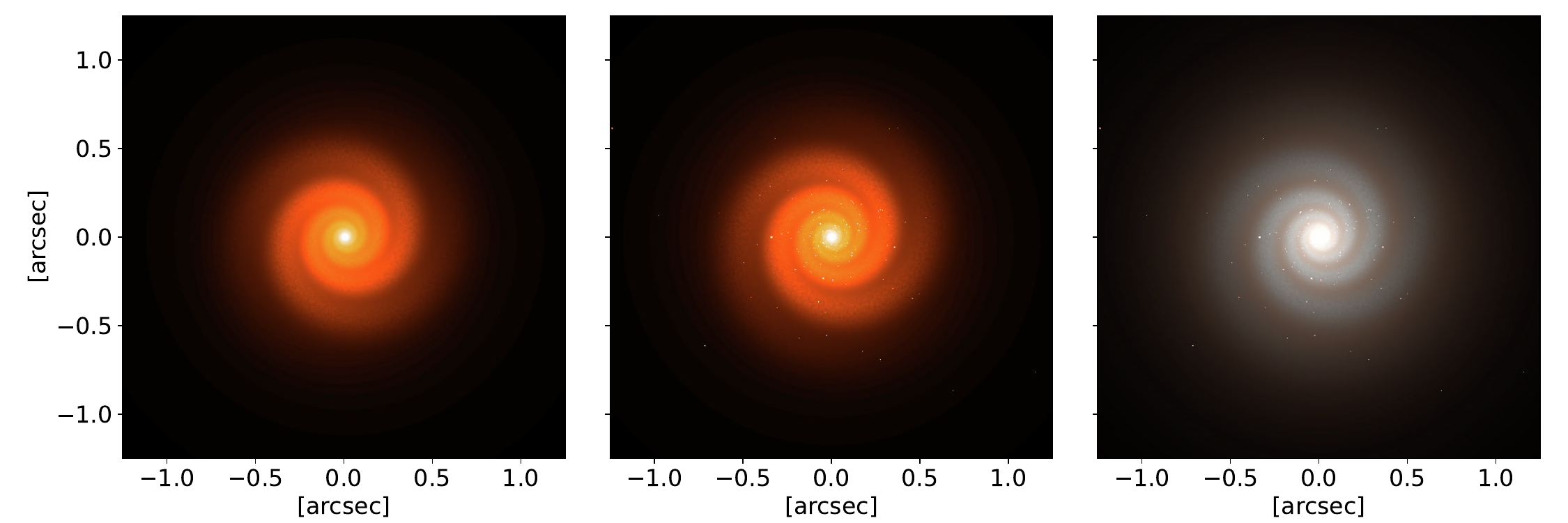}
    \caption{(a) Example of a host galaxy consisting of an exponential disk with two spiral arms and a spherical bulge. Details on the adopted parameters are reported in the text. (b) Model of a host galaxy, made of an exponential disk, a spherical bulge, and two spiral arms, populated with S\'ersic stellar clumps. Details on the adopted parameters are reported in the text. The galaxy center coincides with the center of the field of view. (c) RGB color image of the same source. Here, we assume the bulge is 1 Gyr old, while the disk is 500 Myr old. For the stellar clumps, their ages span in the range 1-500 Myr. In this case, the red, green, and blue channels correspond to effective wavelengths $\lambda_e = 5$, $3$, and $2~\mu m$, respectively (0.70, 0.42 and 0.28 $\rm \mu m$, rest-frame, taking into account an assumed redshift $\rm z=6.145$). These images have an angular resolution of $0.25$ mas/pix, which at a $z \sim 6$ corresponds to a physical scale of approximately 1 pc per pixel. As a result, our clumps, all of which have sizes between 1 and 10 pc, are resolved in the high-resolution images.}
    \label{Unlensed}
\end{figure*}

\subsection{Spiral arms}
\label{sect:spiral}
The disk in the source model outlined in Sect.~\ref{sect:host} has a smooth brightness distribution. Our model allows the addition of a spiral structure to the smooth disk.

To achieve this goal, we follow the method of \citet{Metcalf_2019}, where the surface brightness of a circular disk including $N_a$ spiral arms is described by
\begin{equation}
\begin{gathered}
    I(\beta,\vartheta)_{disk} = I_{0}\exp{\left(-\frac{\beta}{R_{h}}\right)}[1+A \cos(N_a \vartheta + \phi_r)],
\\
    \phi_r = \alpha \log \left(\frac{2\beta}{R_{h}} \right) + \phi_d \;,
\end{gathered}
\label{eq:spiral_arms}
\end{equation}
where $\vartheta$ is a polar angle, $\phi_d$ is the phase angle of the arms, and $\alpha$ describes how the arms are wrapped around the bulge. 

The equations above correspond to a circular disk observed face-on. 
Therefore, in this specific case Eq.~\ref{eq:betatilde} simplifies to
\begin{equation}
    \beta = \sqrt{(\beta_1 - \beta_{s,1})^2 + (\beta_2 - \beta_{s,2})^2}.
\end{equation}

To simulate different viewing angles of the same galaxy, we introduce a new parameter, $\tilde\varphi$, which accounts for the inclination of the disk with respect to the optical axis. 
By incorporating this parameter, Eq.~\ref{eq:betas} becomes
\begin{equation}
\begin{gathered}
    \beta_1' = (\beta_1 - \beta_{s,1})  \cos\varphi + (\beta_2 - \beta_{s,2})  \sin\varphi, 
\\
    \beta_2' = -(\beta_1 - \beta_{s,1})  \sin\varphi + (\beta_2 - \beta_{s,2}) \cos\varphi / \cos\tilde\varphi.
\end{gathered}
\end{equation}

Finally, to make the spiral arms more irregular, we perturb the disk simulating a Perlin noise\footnote{To generate Perlin noise we use the {\tt pnoise2} function from the package {\tt noise}, \href{https://github.com/caseman/noise}{https://github.com/caseman/noise}.}.  

In the first panel of Fig.~\ref{Unlensed}, we show an example of a face-on spiral galaxy ($\tilde\varphi=0$) on the source plane. 
We adopt an axis ratio $q = 1$ for the disk, while the effective radius is $R_{e, {\rm disk}} = 0.18''$.
For the arms we use $A=1.0$, $N_a=2$, $\phi_d=1$, and $\alpha=5.0$. 
The bulge effective radius is $R_{e, {\rm bulge}} = 0.1''$. 
We use Eqs.~\ref{eq:host_gal} and \ref{eq:spiral_arms} to compute the source brightness distribution on a regular grid with $2000\times 2000$ pixels, with an angular resolution of $0.25$ mas/pix. 
The resolution of this grid can be properly chosen according to the purpose of the simulation, as it will be discussed below (see Sect.~\ref{sec:noise}).

In the following Sections, we will use the source galaxy in the left panel of Fig.~\ref{Unlensed} as the host for the stellar clumps. 
In addition, we will calibrate the source flux to mimic the observations of the Cosmic Archipelago arc in galaxy cluster MACS J0416.1-2403; this is a highly magnified z=6 Lyman-$\alpha$ emitter \citep{Vanzella_2017,Vanzella_2019}, hosting massive star clusters, as revealed by recent JWST observations \citep{Messa2025}. 
Since we do not have a proper estimate of the host galaxy stellar mass from SED fitting, we assume a value $M_{*,tot} = 10^8~M_{\odot}$. 
This value is consistent with stellar masses measured for some high-$z$ galaxies \citep[e.g.,][]{Vanzella_2022}, and is therefore a plausible choice for our simulated source.

\subsection{The stellar clumps}
\label{sec:clumps}

This Section describes how we populate the host galaxy with stellar clumps.
We make the following assumptions:
\begin{enumerate}
    \item the clump spatial distribution follows the underlying brightness distribution of the host;
    \item the total mass in clumps is a predetermined fraction of the total mass of the host;
    \item the distribution of clump masses obeys a Schechter-like function \citep[e.g.][]{1976ApJ...203..297S};
    \item the clump size is a growing function of the clump mass. 
\end{enumerate}

While our modelization is extremely flexible, the model parameters discussed below should be set to match observations. 
Unfortunately, the statistical properties of stellar clumps in distant, high-$z$ galaxies are not well constrained \citep[see, e.g.][]{Mestric2022,Claeyssens2025}. 
For this reason, we use observations of local galaxies as {\em mild} references. 
The similarity of these sources to their high-$z$ equivalents is not guaranteed. 
In the future, we plan to investigate the statistical properties of stellar clumps in high-$z$ sources by comparing simulations realized with our tool to real observations.
Such an application can help us understand the level of similarity between clumpy sources in the high- and low-$z$ Universe.

Firstly, we constrain the clump's overall number, considering that only a fraction $Q$ of the host galaxy's total stellar mass goes into star clumps.
This fraction varies with the morphological classification of a galaxy: it is reasonable to assume $Q =$~\([0-0.1]\) for early-type, lenticular galaxies and $Q =$~\([0.2-0.4]\) for late-type galaxies \citep{plezias18}, while even higher fractions can be reached at high-$z$, where extreme gas physical conditions are not rare in star-forming galaxies \citep[][]{Adamo_2020} which can be characterized by high stellar and gas densities \citep[][]{10.1093/mnras/stv300}.
We expect late-type galaxies to be characterized by stronger star formation and host more star-forming regions than early-type galaxies.
For the simulation of the source at $z = 6.145$, we can assume extreme star formation conditions by using $Q = 0.8$. 

We assume that the clump stellar masses follow a distribution which is a generalization of a Schechter mass function,
\begin{equation}
f(x) = x^{\beta} \exp{\left[-{\delta \left( \frac{x}{x_{cut}} \right)^{\gamma}}\right]} \;,
\label{lum_func}
\end{equation}
which combines a power-law behavior with slope $\beta$ at low masses with an exponential cut-off above a characteristic scale $x_{cut}$. 
In the previous equation, $x$ is the clump mass in solar masses and  $\delta$ and $\gamma$ are parameters defining the shape of the mass function in the high-mass tail. 
This parameterization of the clump mass function is borrowed from cosmological hydrodynamical simulations of galaxy formation and evolution \citep[e.g.][]{2010RSPTA.368..867L}. 
It can be adapted to observations of local galaxies, which are consistent with setting $\beta = -2.0$, $\delta = 1.0$, and $\gamma =1.0$ \citep[e.g.][]{Krumholz2019,Adamo2020}. 

\begin{figure}[t]
    \centering    
    \includegraphics[width=\linewidth]{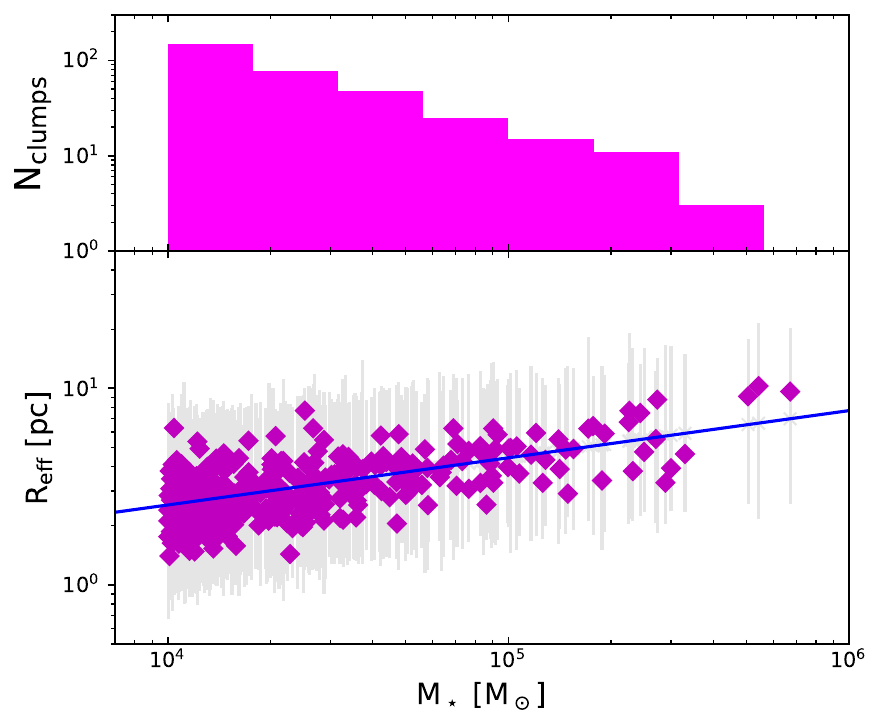}
    \caption{Upper panel: Magenta
    histogram showhing the stellar mass function of the star clumps populating our simulated galaxy. Lower panel: Clump sizes (in units of parsec) as a function of the clump stellar mass. Clumps are represented by magenta points. The underlying average size-mass relation given by Eq. \ref{r_eff} \citep[][]{brown2021} is shown in a blue solid line. At a given stellar mass, we assume the size distribution follows a lognormal distribution with a standard deviation of $\sigma = 0.3$.}
    \label{fig:cl_dim}
\end{figure}

We sample the mass function in the range between $x_{min}=~10^{4}~M_{\odot}$ and $x_{max} = 10^{7}~M_{\odot}$. 
Furthermore, we define $x_{cut}= 5 \times 10^6~M_{\odot}$, to limit the number of very massive stellar clumps, considering that the total galaxy mass is assumed to be $10^8 M_{\odot}.$
The sampling continues until we saturate the mass budget in stellar clumps, as defined by the parameter $Q$. 
To conserve the total galaxy mass, we subtract the sum of the clump masses from the host mass.
The upper panel of Fig.~\ref{fig:cl_dim} shows the stellar mass function of the star clumps resulting from this sampling.

As anticipated, we assume that the clump spatial distribution follows the surface brightness distribution of the host galaxy. 
This assumption is reasonably well supported by observations of local galaxies \citep[see e.g.][]{Adamo2020}. 
Thus, we draw the clump positions in the host by sampling the source pixel surface brightness distribution.

We model the stellar clumps using S\'ersic profiles with $n=0.5$, assuming they have elliptical shapes with axis ratios uniformly distributed in the range [0.3,1). 
Their position angles are also randomly chosen in the range [0, $\pi$).
To determine their sizes, we use the relation between effective radius and clump mass measured from YSCs in the Local Universe by \cite{brown2021}:
\begin{equation}
R_{e} = 2.55  \left( \frac{M}{10^4 M_{\odot}} \right)^{0.24}\;.
\label{r_eff}
\end{equation}
\citet{Claeyssens2023} proposed a different mass-size relation for clumps with sizes $\geq 10$ pc in high-$z$ ($z>~1$) galaxies\footnote{Being based on larger scales than individual clusters, the \citet{Claeyssens2023} relation characterize systems with lower densities than the \citet{brown2021} ones.}. Since we want to simulate objects with sizes down to $\sim$pc scales, we prefer to rely on the \citet{brown2021} relation.
Hence, as shown in the lower panel of Fig.~\ref{fig:cl_dim}, our clumps can span a range of effective radii from $\sim10$ pc down to $\sim1$ pc. 
Moreover, at a given stellar mass, we assume that the size distribution follows a lognormal distribution with a standard deviation of $\sigma = 0.3$.

The second panel of Fig.~\ref{Unlensed} shows the same galaxy in the first panel but after including the stellar clumps as discussed in this Section. 
Further examples of possible morphological variations of the source galaxy, as well as variation in the cluster mass fraction, are presented in Appendix \ref{app:shape}, highlighting the flexibility of our pipeline. 

\begin{figure}[b!]
    \centering    
    \includegraphics[width=\linewidth]{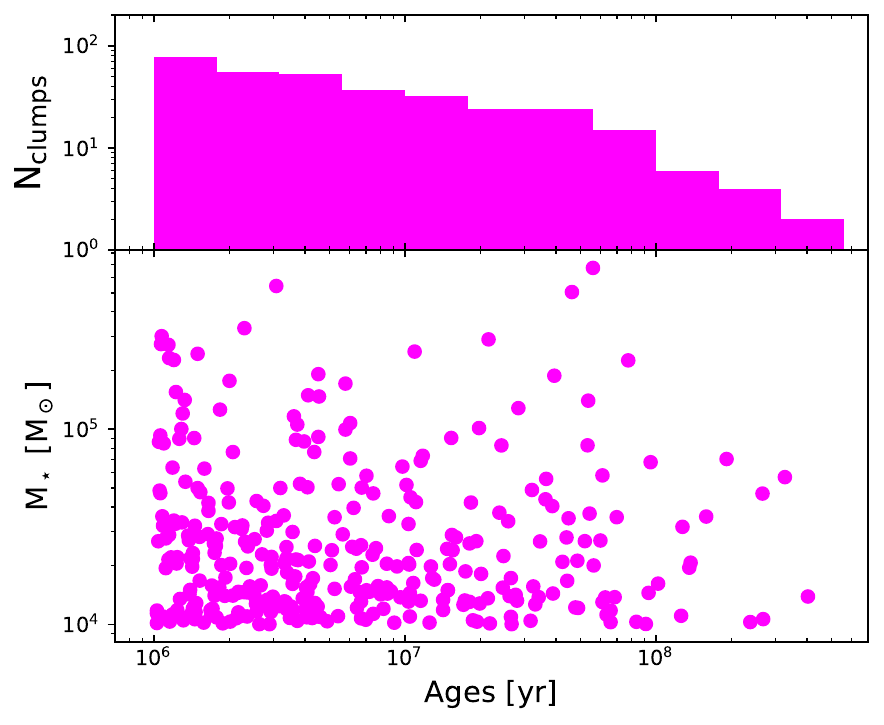}
    \caption{Upper panel: Histogram of the clump's ages. Lower panel: Clump stellar masses as a function of their ages in years.}
    \label{fig: cl_ages}
\end{figure}

\subsection{Simulating colors}
\label{sec:colors}
Since we want to produce multiband observations of the clumpy galaxy, we need to assign Spectral Energy Distributions (SEDs) to the host and clumps.
To derive the SEDs of each component and, therefore, to produce the expected magnitudes for different instruments and filters, we use stellar population synthesis models. 
In particular, we use the Yggdrasil population synthesis code by \citet{Zackrisson_2011}, designed for modeling the SEDs of high-$z$ galaxies~\footnote{\url{https://www.astro.uu.se/~ez/yggdrasil/yggdrasil.html}}. 
Yggdrasil predicts the SEDs of integrated stellar populations across a wide range of metallicities, from zero to supersolar, accounting also for nebular emission and dust extinction.
The software allows the SEDs to be computed directly or the magnitudes to be computed in any photometric band for a chosen instrument.
Among the input parameters required by Yggdrasil to compute the SEDs are the metallicity ($Z$) and the Initial Mass Function (IMF) for the stellar population; for the examples of the current work, we always assume $Z = 0.004$, with a Kroupa IMF \citep[e.g.][]{2001MNRAS.322..231K}. 
One further assumption is the Star Formation History (SFH) of the system. 
In particular, the Star Formation Rate (SFR) can be instantaneous, which produces a single-age stellar population, or constant over a specific period of time. 
For our simulation, we assume for both components of the host galaxy a constant SFR for $100$~Myr, while, for the stellar clumps, we assume a constant SFR for $10$~Myr. 
We assume the bulge to be $1$~Gyr old, while the spiral arms are assumed to be younger, with an age of $500$~Myr. 
Regarding the stellar clumps, we define their ages by sampling a power-law distribution with slope $\alpha = -0.5$ and spanning between $1$~Myr and $500$~Myr. 
This is in line with cluster studies in nearby galaxies \citep[see e.g.,][]{Krumholz2019}.
The upper panel of Fig.~\ref{fig: cl_ages} shows the distribution of the clump's ages, while the lower panel shows the clump masses as a function of their ages.
We point out that different assumptions can be made on the age and SFHs of the clusters and their host; also the use of other stellar population synthesis models can be easily implemented within the pipeline.

The SEDs obtained from all these ingredients must be redshifted according to the source redshift $z_{s}$. 
Then, they need to be normalized to the correct flux.
Along with the wavelength and flux $F_{Ygg}$ data points, the software provides also the stellar mass (in solar masses) $M_{Ygg}$ of the stellar population associated with that SED. The normalization is thus achieved by assuming a constant mass-to-light ratio:
\begin{equation}
    F_{norm} = F_{Ygg} \frac{M}{M_{Ygg}},
\end{equation}
where $M$ is the mass of our object.

In the right panel of Fig.~\ref{Unlensed}, we show the RGB colored image of our source galaxy, built using the monochromatic flux from simulated images at the three observed wavelengths: $\rm 5~\mu m$ (red channel), $3~\mu m$ (green channel) and $2~\mu m$ (blue channel).
The intrinsic, rest-frame wavelengths are approximately $\rm 0.70~\mu m$, $\rm 0.42~\mu m$, and $\rm 0.28~\mu m$, respectively, taking into account the source redshift of $z = 6.145$.

\section{Simulating lensing effects}
\label{sec:lensing}

\begin{figure}[t]
    \centering    
    \includegraphics[width=\linewidth]{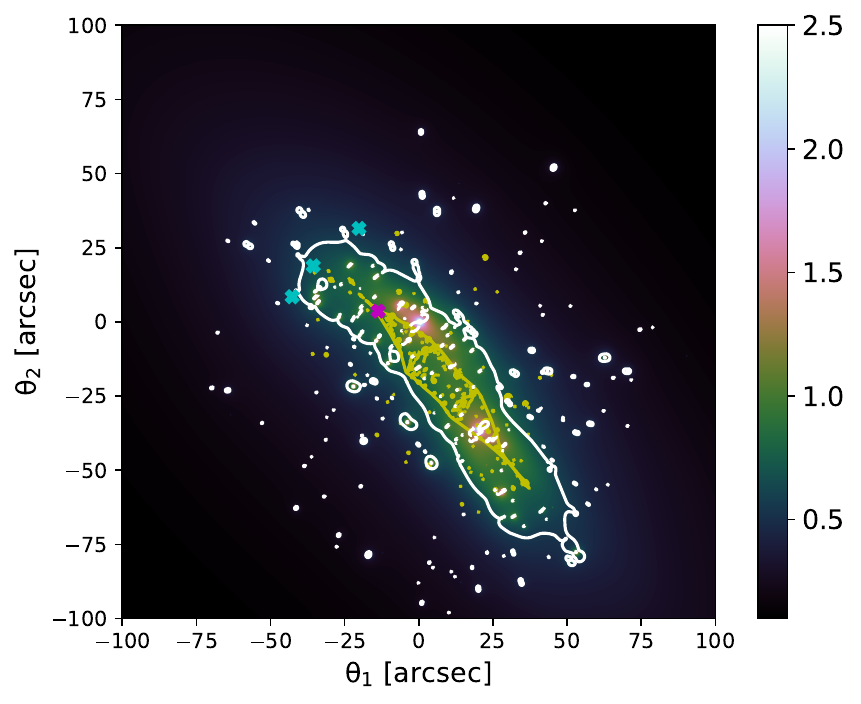}
    \caption{Mass map of galaxy cluster MACS~J0416.1-2403 \citep{Bergamini_2023} at $z = 0.396$ (RA: 64.038142 deg, Dec: -24.067472 deg). This model is generated with the software \texttt{Lenstool}. The yellow lines represent the caustics of a source at $z = 6.145$, while the white lines are the corresponding critical lines on the lens plane. The magenta marker indicates the position of our source ($\beta_1 = -13.72$ arcsec and $\beta_2 = 3.62$ arcsec), and the three cyan markers show the corresponding multiple images on the lens plane: a stretched arc at the position  $\theta_1 = -35.57$ and $\theta_2 = 18.85$, and two counter-images in $\theta_1 = -42.66$ and $\theta_2 = 8.46$ and $\theta_1 = -20.16$ and $\theta_2 = 31.53$), respectively. In this work, we focus on the gravitational arc.}
    \label{fig: lens_models}
\end{figure}

Our pipeline allows us to include lensing effects by providing deflection angle maps. 
These maps can be derived differently (e.g. using analytical lens models or ray tracing through numerical simulations). In this work, we use maps obtained by modeling existing galaxy clusters with the public software \texttt{Lenstool} \citep[see, e.g.][]{kneib96, jullo07, jullo09}. 
More precisely, the examples in this work use the reconstruction of galaxy cluster MACS~J0416.1-2403 at redshift $z=0.396$. 
The details of the mass modeling can be found in several papers \citep{bergamini19, bergamini20, bergamini23}. 
The mass map for this cluster is shown in Fig.~\ref{fig: lens_models}, where the white lines represent the cluster critical lines, that is, the contours corresponding to an infinite magnification on the lens plane, for a source at redshift $z_s=6.145$. 
By mapping the critical lines onto the source plane, we compute the lens caustics, represented in yellow in the same Figure.  
From the position of the caustics, we can identify the points on the source plane where sources can be placed to maximize their magnification and distortion.
We select the position on the source plane at $z_s=6.145$ for our simulated galaxy that reproduces the same configuration and distortion of the Cosmic Archipelago in MACS~J0416.1-2403 \citep[][]{Messa2025}. This configuration features an elongated arc and two counter-images, which, however, are not later included in our simulations. 
The chosen source position lies inside the cusp of the cluster caustic, ensuring that the lensing effect produces three multiple images with high tangential magnification, as they are located near the tangential critical lines. In Fig.~\ref{fig: lens_models}, the magenta point marks the selected source position, while the cyan crosses indicate the corresponding multiple images positions.
Note that 1) the lensed images are stretched and have a spatial extension much larger than the intrinsic source size due to lens magnification; and 2) the stellar clumps in the image are also lensed and multiply imaged inside the arcs, as they follow the same distortion pattern as the overall galaxy.

\begin{figure*}[h!]
      \centering
		\includegraphics[width=0.95\textwidth]{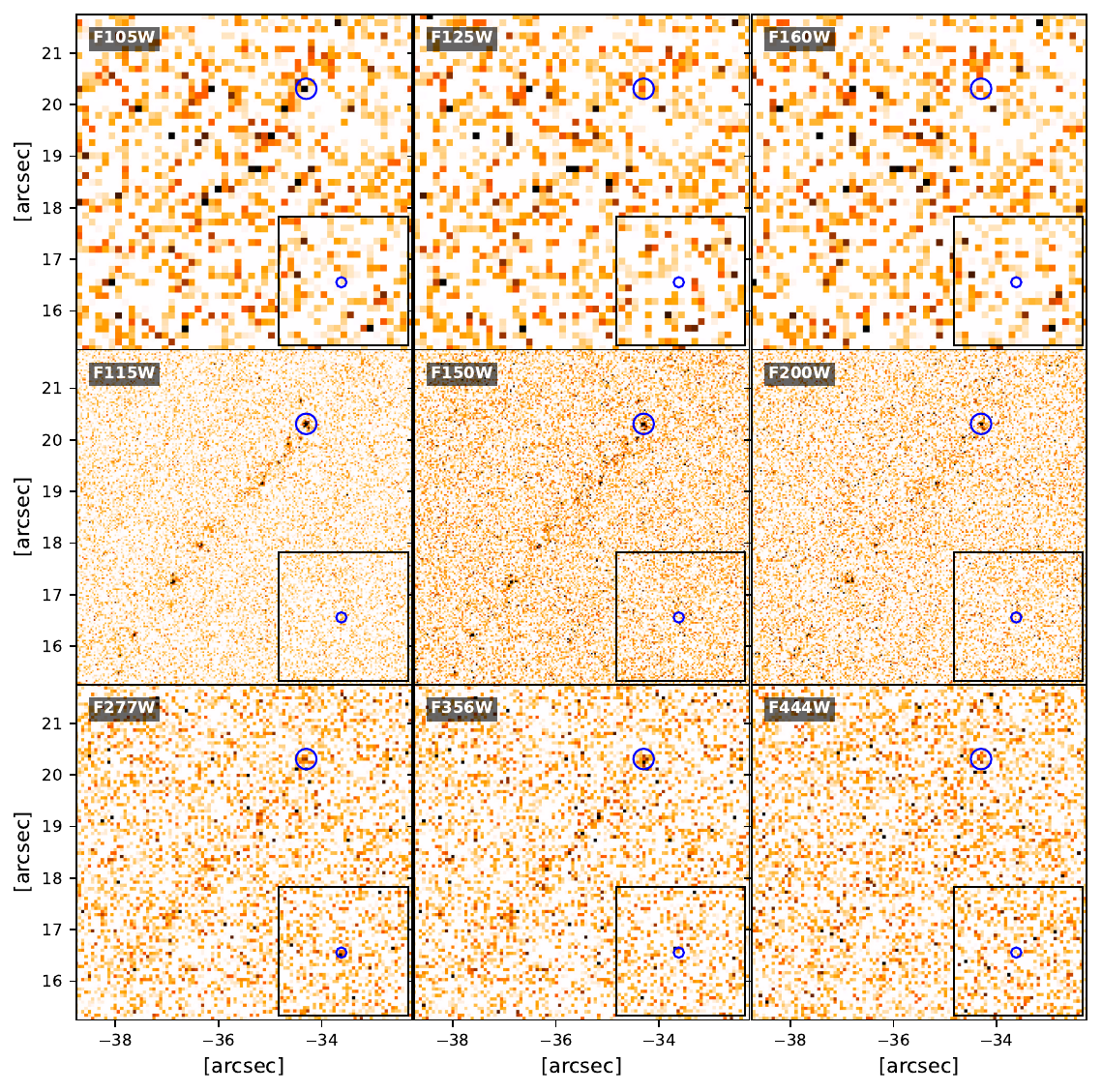}
       \caption{Simulated observations using HST WFC3 (the top three panels, for F105W, F125W, and F160W) and JWST NIRCam (the remaining panels for F115W, F150W, F200W, F277W, F356W, and F444W). In each subfigure, we show the same source at $z = 6.145$ lensed by MACS J0416.1-2403. 
       Each panel uses its own flux range to optimize visibility of the source structure. The typical background level and exposure time for each filter are listed in Table \ref{noise_HST_WFC3} and \ref{noise_JWST}.
       Further details are reported in the text. The black squares in the bottom right corner of every subfigure show how the source galaxy appears when the gravitational lensing effect is not included. The unlensed and lensed sources are represented using the same pixel scale. A blue circle highlights the same clump in both the lensed and unlensed images, showing its corresponding position in the lens and source planes.}
	\label{simul}
\end{figure*}

\section{Inclusion of observational noises}
\label{sec:noise}

This Section illustrates the main steps to mimic observations with specific instruments and add the appropriate observational noise to the previously generated noise-free images. 

For example, we simulate observations of the clumpy galaxies with the Hubble Space Telescope (HST) and James Webb Space Telescope (JWST). 
The procedure we follow is the same for both of them and consists of the following steps:
\begin{enumerate}
\item we draw the proper surface brightness distribution on a pixel grid resembling the resolution of the detector;
\item we convert the source surface brightness into units of counts $s^{-1}$ pixel$^{-1}$;
\item we convolve the resulting image with the instrument Point-Spread-Function (PSF);
\item we add the sky background;
\item we add photon noise.
\end{enumerate}
We do not consider other instrumental effects such as Charge Transfer Efficiency or Brighter Fatter effects, nor do we simulate cosmic rays, radiation damage, hot pixels, etc. 
We assume that these effects have already been corrected in the images we produce.

We begin by assigning a stellar mass to the source we want to simulate, based on observations of the Cosmic Archipelago, in order to ensure that the simulated image is comparable to the observed one. As mentioned in Sect.~\ref{sect:spiral}, we adopt a total stellar mass of $M_{*,tot} = 10^8~M_{\odot}$. 
While we currently lack a robust estimate of the Cosmic Archipelago’s stellar mass from SED fitting, this value is consistent with the stellar masses measured for some high-$z$ galaxies \citep[e.g.,][]{Vanzella_2022}, and is therefore a plausible assumption for our simulation.
In this work, we split the total stellar mass into the disk and the bulge. 
Moreover, the star clump mass is a fraction of the disk's.
In particular, we refer to a disk of $M_{*,disk} = 9.0 \times 10^{7} M_{\odot}$, a bulge of $M_{*,bul} = 1.0 \times 10^{7} M_{\odot}$, and 329 stellar clumps with a total mass of  $M_{*,cl} = 7.2 \times 10^{7} M_{\odot}$, corresponding to a fraction $Q = 80\%$ of $M_{*,disk}$. We point out that the disk mass includes the mass in clumps.

At first, our tool constructs the source with an angular resolution significantly higher than the telescope detector's chosen for the simulation. In this example, we adopt 0.25 mas/pix instead of the resolution of the Wide Field Camera 3 (WFC3) onboard HST (0.13 arcsec/pix) or the Near Infrared Camera (NIRCam) onboard JWST (short wavelength channel: 0.031 arcsec/pix, long wavelength channel: 0.063 arcsec/pix). 
This choice is driven by the need to ensure that the clumps within the source are well sampled before simulating lensing effects and the observation with a specific instrument. 
If the source were not constructed at such a high resolution, the lensing effect would be applied to intrinsically unresolved clumps, causing them to appear as point-like sources rather than revealing their actual extended structure.
Consequently, unresolved clumps would not exhibit the stretched features characteristic of gravitational arcs, but would instead appear as compact, magnified points in the lens plane.
While a fine resolution of 0.25 mas/pix is necessary on the source plane to resolve such high-$z$ clumps, we use a coarser resolution of 2.5 mas/pix on the lens plane, which still preserves the morphological details of lensed features in the final simulated images.

Using the instrument Zero-Point (ZP), we convert the magnitudes obtained in Sect.~\ref{sec:colors} into a total number of counts (or data numbers, DN) $s^{-1}$ registered by the detector. 
The ZP of an instrument, by definition, is the magnitude of an object that produces one DN per second. 
The magnitude of an object producing DN counts in an observation of length $t_{exp}$ is, therefore:
\begin{equation}
m_{\rm AB} = - 2.5\log_{10}\left(\frac{\rm DN}{t_{exp}}\right) + {\rm ZP}_{\rm AB} \;.
\end{equation}
To obtain DN $s^{-1}$, we invert the equation above, using the exposure time $t_{exp}$ from the real observation that serves as a reference for our simulations.
We then scale the brightness distribution of the source so that the total number of DN $s^{-1}$ matches the value corresponding to the desired input magnitude.

In the next step, we consider the PSF of the chosen instrument. 
We apply it to the image of the clumpy galaxy in units of DN $s^{-1}$ pixel$^{-1}$ via a two-dimensional convolution:
\begin{equation}
\tilde{I}(\theta_1,\theta_2) = \int \int d\theta'_1 d\theta'_2 I(\theta'_1,\theta'_2) {\rm PSF}(\theta_1-\theta'_1,\theta_2-\theta'_2) \;.
\end{equation}

The sky background (bkg hereafter) in an astronomical observation depends on several factors, including the direction in which the observation is carried out. 
In the examples shown in this paper, we set the background level according to real observations of the sources we wish to simulate. 
More precisely, we place 50 circular regions over the darkest areas of real images, particularly avoiding every source. The mean bkg flux associated with each pixel can be obtained from these regions.

Finally, the photon noise is computed on the pixel grid, assuming it follows a Poisson distribution with variance given by the sum of the source and background counts. The simulated raw observations are obtained by summing the noise maps and the PSF-convolved source images. More details about this procedure can be found in \citet{Meneghetti2008} and \citet{plezias18}, for example.

\section{Preparing the virtual observations}
\label{sec:virtualobs}
As anticipated earlier, to illustrate how our simulation software works, we chose to focus on a source inspired by the gravitationally lensed source Cosmic Archipelago. 
This source is particularly interesting for the stellar clumps it contains. 
We intend not to reproduce the Cosmic Archipelago exactly, but to simulate a similar source size and lensing configuration. 
Hereafter, we detail the parameters required to model the source as a S\'ersic galaxy, specifying those related to the bulge and the disk components. The full list of keywords defined in our pipeline to model the source and to set up the virtual observations are reported in Appendix \ref{app:keywords}.

\begin{itemize}
    \item Bulge: S\'ersic index $n=4$, effective radius $R_e=0.1''$, axis ratio $q=1$, position angle $\phi = \pi/8$; 
    \item Disk: S\'ersic index $n=1$, effective radius $R_e=0.18''$, axis ratio $q=1.0$, position angle $\phi = \pi/8$;
\end{itemize}
Both components are centered at source positions $\beta_{s,1} = -13.716581''$ and $\beta_{s,2} = 3.6176634''$, and set at redshift $z_s=6.145$. 
Regarding the spiral arms, we use these parameters: scale height $R_h = 0.1''$, normalization $A=1.0$, number of arms $N_a = 4$, linking angle $\phi_d = 1.0$, wrapping factor $\alpha = 10.0$, tilting angle $\phi = \frac{3}{4}\pi$. 
At this stage, the simulation produces an ideal high-resolution image of the source galaxy, without any instrumental effects such as PSF convolution, exposure time, or noise. 
This image is shown in the central panel of Fig.~\ref{Unlensed}.

In this example, we simulate observations with two cameras onboard the HST and JWST, namely WFC3 in the IR channel and NIRCam.
Different pixel scales and PSFs characterize these instruments. 
Some details are given below.

\paragraph{HST WFC3/IR} We simulate observations through the IR Channel of the imager WFC3, onboard HST.
The native pixel size of this detector is $0.13''$/pix. 
We consider the bands F105W, F125W, and F160W.
The PSF models used in our simulations are obtained with the public PSF modeling tool \texttt{Tiny Tim} \citep{tinytim}. 

\begin{figure*}[h!]
    \centering   
    \includegraphics[width=0.9\linewidth]{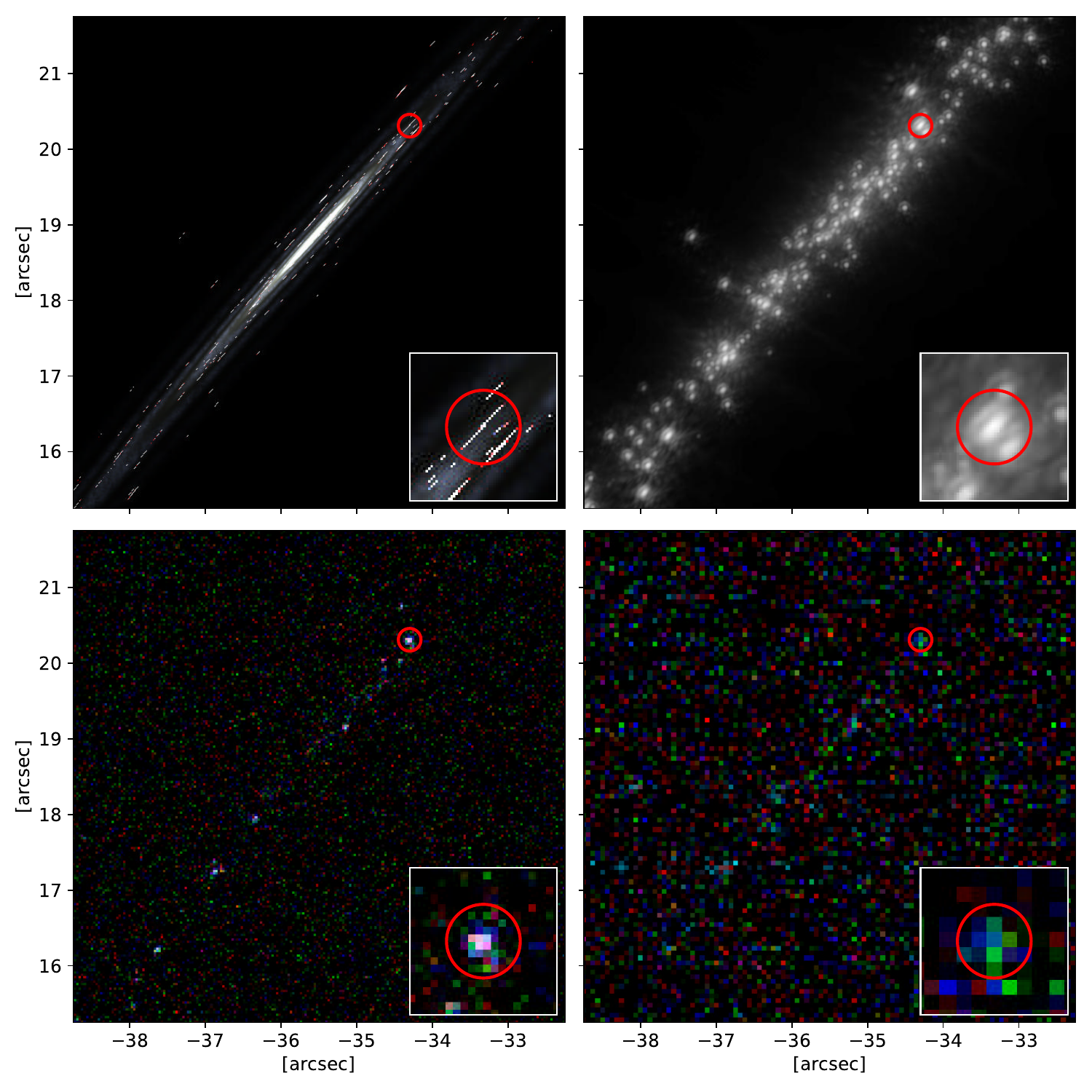}
    \caption{RGB simulation of the Cosmic Archipelago ($z = 6.145$), which is in MACS J0416.1-2403. This is the same source shown in the right panel of Fig. \ref{Unlensed}, but including strong lensing effects by MACS J0416.1-2403. The unlensed source is assumed to be at $z = 6.145$ and distorted into a gravitational arc if placed near the cluster tangential caustic. Top left: a high-resolution RGB simulation ($2.5$ mas/pix), where the red, green, and blue channels correspond to wavelengths of $\lambda_e =$ 5, 3, and 2 $\mu$m, respectively; top right: the same high-resolution image, but monochromatic, constructed using the magnitudes of the host and the clumps in the JWST/F115W filter and convolved with the JWST/F115W PSF to illustrate the effect of instrumental resolution; bottom left: an RGB image created by combining three JWST NIRCam short wavelength channels (red F200W, green F150W, and blue F115W), including realistic noise and sky background; bottom right: an RGB image created by combining  three JWST NIRCam long wavelength channels (red F444W, green F356W, and blue F277W), including realistic noise and sky background. The background level and effective exposure time are based on real JWST observations of MACS J0416.1-2403 and are listed in Table \ref{noise_JWST}. The red circle in each image highlights an example of a clump, illustrating how the instrument's resolution is a limiting factor: while clumps appear as distinct structures in the HR image (see the zoom-in region shown at the bottom right of the figure), they become blended together in the JWST observations. For each image, the origin of the coordinate system coincides with the center of the cluster, located at (RA: 64.038142 deg, Dec: –24.067472 deg).}
	\label{fig:RGB_M0416}
\end{figure*}

To define the sky background and exposure time, we use the observations of MACS~J0416.1-2403 from the CLASH program. 
While $t_{exp}$ is explicitly stated in the data header, the bkg level is estimated following the same approach described in Sect. \ref{sec:noise}. Specifically, we analyze the CLASH files by placing 50 circular regions over the darkest areas, carefully avoiding all luminous sources.
The mean $bkg$ flux associated with each pixel can be obtained from these regions.
The $t_{exp}$ values are listed in the first row of Table~\ref{noise_HST_WFC3}, while the corresponding $bkg$ are reported in the second row of the same Table.

\begin{table}[h!]
    \centering
    \captionsetup{position=above}
    \caption{Values of $t_{exp}$ (first row) and $bkg$ fluxes in (second row) for HST WFC3/IR in the filters F105W, F125W, and F160W.}
    \label{noise_HST_WFC3}
    \resizebox{\columnwidth}{!}{%
    \begin{tabular}{||c || c c c||}
     \hline
         & F105W & F125W & F160W \\
        \hline\hline
         $t_{exp}$ [s] & 5511.747 & 5511.747 & 5511.747 \\ 
         \hline
         $bkg$ [counts/s/pix] & 0.694285 & 0.547766 & 0.5331 \\
         \hline
    \end{tabular}
    }
\end{table}

\paragraph{JWST NIRCam} We simulate observations in the imaging mode of the JWST NIRCam through the short-wavelength channel (0.6-2.3 $\mu$m) in the bands F115W, F150W, and F200W and through the long-wavelength channel (2.4–5.0 $\mu$m) in the bands F277W, F356W and F444W.
In this case, the pixel scale is $0.031''$/pix for the short wavelength channel and $0.063''$/pix for the long wavelength one. 
The PSF models for the short-wavelength channel are taken from \cite{plezias18}\footnote{\url{https://www.stsci.edu/jwst/science-planning/proposal-planning-toolbox/psf-simulation-tool}}, who generated them using the tool {\sc WebbPSF}, while those for the long-wavelength channel are directly generated from the same tool. 
We analyze the data of MACS J0416.1-2403 from the Cycle 1 GTO program (PI Windhorst, \citet{Windhorst_2022}) and define $t_{exp}$ and $bkg$ following the same method used for HST WFC3/IR. 
These values are listed in Table~\ref{noise_JWST}.

\begin{table}[h!]
    \centering
    \centering
    \captionsetup{position=above}
    \caption{As Table~\ref{noise_HST_WFC3}, but for JWST NIRCam (F115W, F150W, F200W, F277W, F356W, and F444W).}
    \label{noise_JWST}
    \resizebox{\columnwidth}{!}{%
    \begin{tabular}{||c || c c c ||}
     \hline
         & F115W & F150W & F200W \\
        \hline\hline
        $t_{exp} [s]$ & 3349.872 & 2920.4 & 2920.4 \\ 
         \hline
         $bkg$ [counts/s/pix] & 0.207835 & 0.20776 & 0.160606 \\
        \hline\hline
         & F277W & F356W & F444W \\
        \hline\hline
        $t_{exp}$ [s]& 2920.4 & 2920.4 & 3779.34 \\ 
         \hline
         $bkg$ [counts/s/pix] & 0.0948052 & 0.0935602 & 0.2402224 \\
        \hline
    \end{tabular}
    }
    \end{table}

\section{Results}
\label{sec:disc}

Figure~\ref{simul} shows the observations simulated with our pipeline. 
In particular, in the first row, we can see the images obtained simulating HST WFC3/IR observations, and in the second and third rows, those coming from JWST NIRCam. 
A preliminary visual inspection of our simulations reveals the detection of only a few clumps, consistently with what is observed in real images of the Cosmic Archipelago \citep[][]{Messa2025}. The comparison between HST WFC3/IR and the short-wavelength (SW) filters of JWST NIRCam highlights how, despite the shorter exposure times, the better resolution and sensitivity of JWST allow the detection of some (observationally) faint clusters. Nevertheless, we note how sources that are at the detection limit in the SW NIRCam filters may be lost again in the long-wavelength (LW) ones.
As a demonstration, the clump highlighted in Fig.~\ref{simul} is undetectable in the HST simulations (with estimated signal-to-noise ratios S/N $\sim 0.7$ in F105W, $\sim 0.4$ in F125W, and $\sim 0.3$ in F160W), while it remains clearly visible in all short-wavelength NIRCam images (reaching S/N $\sim 14$ in F115W, $\sim 8$ in F150W, and $\sim 10$ in F200W). In contrast, its significance decreases in the long-wavelength NIRCam bands (S/N $\sim 4$ in F277W, $\sim 5$ in F356W, and $\sim 1.5$ in F444W). If we consider S/N $ < 3$ as very low signal-to-noise ratios \citep[][]{Messa2025}, it is clear that in the HST simulations the clump is undetectable, in the JWST LW bands it lies close to the detection limit, while it is clearly detectable in the JWST SW bands.

To assess the validity of our tool and enable a meaningful comparison with real observations, we compared the limit magnitude of our simulated images with that of the real images.
By placing 100 circular apertures over dark regions of the simulated observation, we can define a numerical distribution of the background counts associated to each region. In the case of JWST/NIRCam F200W band, the standard deviation of those apertures corresponds to an AB limit magnitude $\text{mag}_{\text{lim}} = 27.38$.
By repeating the same steps on the real data, we get the comparable limiting magnitude $\text{mag}_{\text{lim}} = 27.42$, supporting the validity of our methodology. 
From a qualitative comparison to the compact rest-UV sources seen in the Cosmic Archipelago (e.g. Fig.~2 in \citealp{Messa2025}), it is clear that the clusters we are simulating are fainter than their `real' counterpart; indeed, while all our clusters have masses $\rm \leq5.8\cdot10^5~M_\odot$ (see Fig.~\ref{fig:cl_dim}), the lowest mass reported in \citet{Messa2025} is $\rm 10^6~\ M_\odot$.  

\begin{figure*}[b]
      \centering		\includegraphics[width=0.95\textwidth]{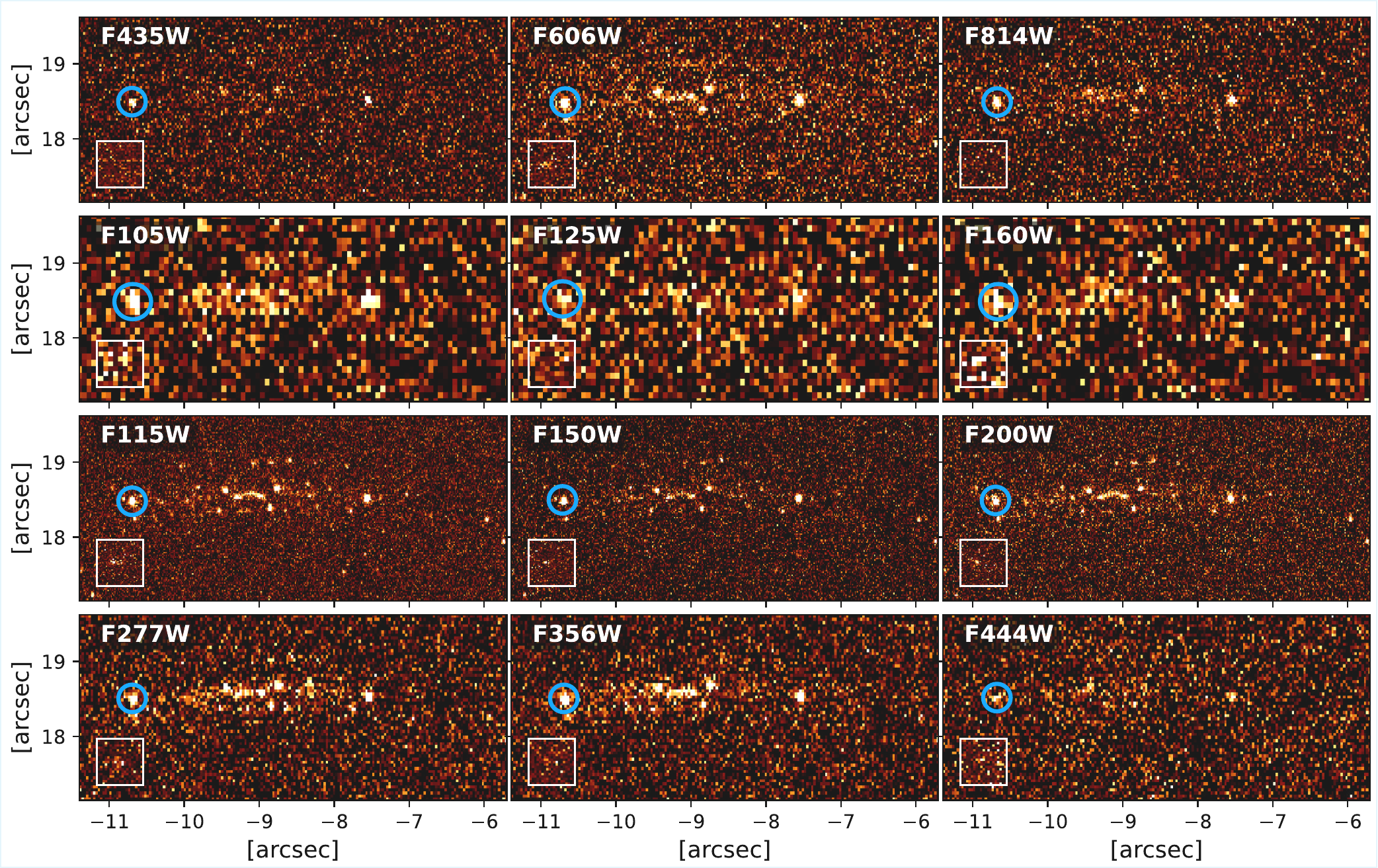}
       \caption{Simulated observations of the System 3 at $z = 3.98$ lensed by Abell 2744, using the HST ACS filters F435W, F606W, and F814W; the HST WFC3/IR filters F105W, F125W, and F160W; and the JWST NIRCam filters F115W, F150W, F200W, F277W, F356W, and F444W. Each panel uses its own flux range to optimize visibility of the source structure. The typical background level and exposure time for each filter are listed in Table~\ref{noise_Abell2744}. Further details, including the S/N analysis of the clump marked with a blue circle, are reported in the text. are reported in the text. The white squares in the bottom left corner of every panel show how the source galaxy appears when the gravitational lensing effect is not included. The unlensed and lensed sources are represented using the same pixel scale.}
	\label{simul_A2744}
\end{figure*}

In Fig.~\ref{fig:RGB_M0416}, we show two panchromatic views of the simulated Cosmic Archipelago. 
In particular, in the two bottom panels, we show a color composite image created by combining three JWST NIRCam/SW channels (red F200W, green F150W, and blue F115W); and three JWST NIRCam/LW channels (red F444W, green F356W, and blue F277W). It is worth noticing that these images reproduce the same zoom-in over the Cosmic Archipelago seen with high-resolution in the first panel of  Fig.~\ref{fig:RGB_M0416}.
We use the red circle to trace back the position of the observed brightest clump to the full-resolution lensed image (top-left panel). Several individual clusters are imaged within the circle, as further revealed by the convolution to the F115W PSF (the top-right panel), providing an intermediate view of the lensed galaxy, prior to the inclusion of instrumental noise and sky background. In our simulated system, the fact that only a `single' clump is detected within the red circle, is caused by one of the clumps being much brighter than the others. This highlights how faint clumps, or regions with low signal, can conceal a much more complex underlying structure. A more quantitative analysis will be carried out in future works.

\section{Abell 2744}
\label{sec:a2744}

\begin{figure*}[!t]
      \centering		\includegraphics[width=0.95\linewidth]{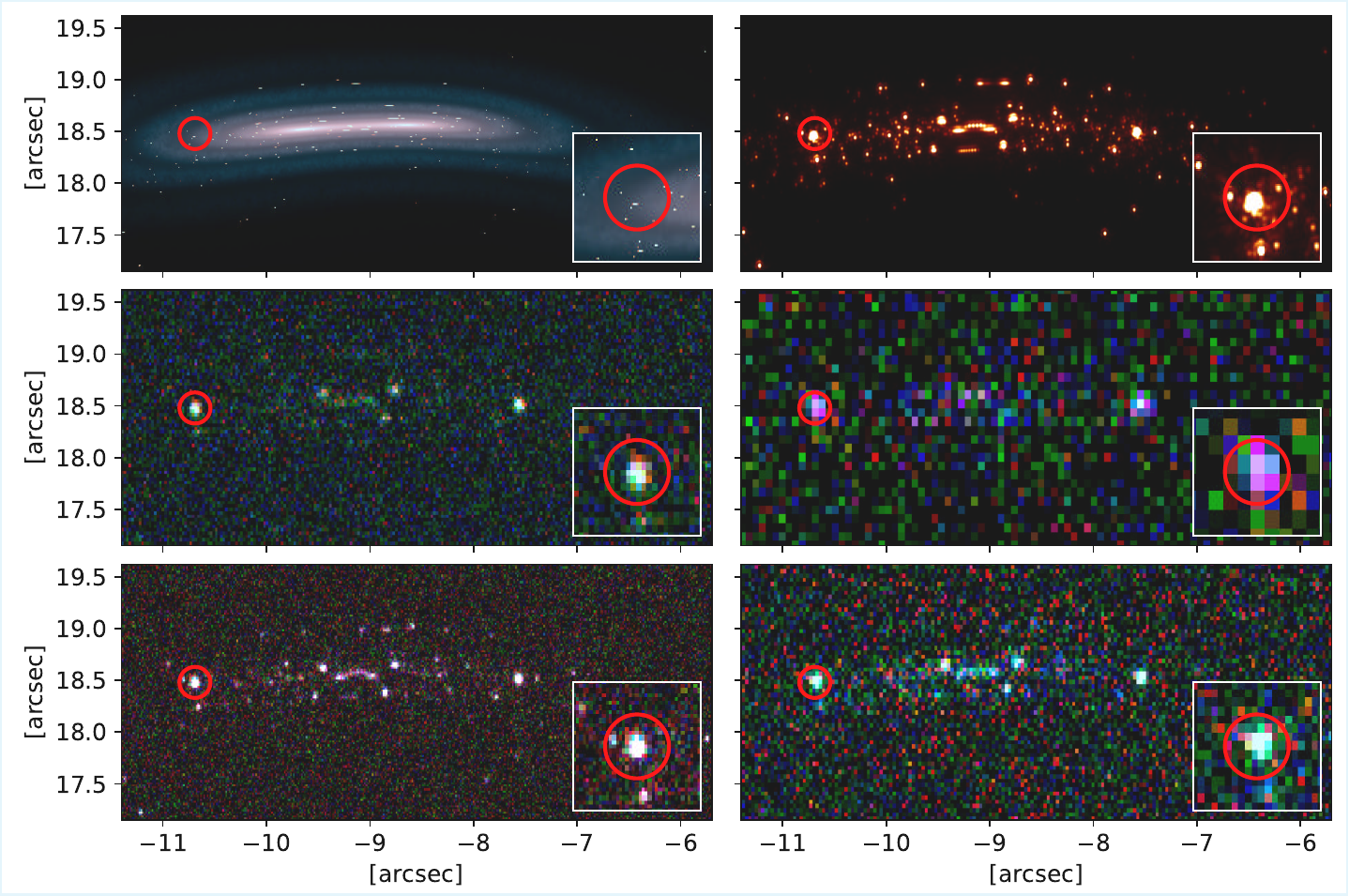}
       \caption{RGB simulations of the System 3 at $z = 3.98$ in Abell 2744. Top left: a high-resolution RGB simulation (2.5 mas/pix), where the red, green, and blue channels correspond to wavelengths of $\lambda_e = $ 5, 3, and 2 $\mu$m, respectively; top right: the same high-resolution image, but monocromathic, constructed using the magnitudes of the host and the clumps in the JWST/F115W filter and convolved with the JWST/F115W PSF to illustrate the effect of instrumental resolution; middle left: an RGB image created by combining three HST ACS bands (red F814W, green F606W, and blue F435W), including realistic noise and sky background; middle right: an RGB image created by combining three HST WFC3/IR bands (red F160W, green F125W, and blue F105W), including realistic noise and sky background; bottom left: an RGB image created by combining three JWST NIRCam/SW channels (red F200W, green F150W, and blue F115W), including realistic noise and sky background; bottom right: an RGB image created by combining three JWST NIRCam/LW channels (red F444W, green F356W, and blue F277W), including realistic noise and sky background. The background level and effective exposure time are based on the data of Abell 2744 from real HST and JWST observations and are listed in Table~\ref{noise_Abell2744}.
       The red circle in each image highlights an example of a clump, illustrating how the instrument’s resolution is a limiting factor: while clumps appear as distinct structures in the HR image (see the zoom-in region shown at the bottom right of the figure), they become blended together in the JWST and HST observations. For each image, the origin of the coordinate system coincides with the center of the cluster, located at (RA: 3.586257 deg, Dec: –30.400172 deg).}
	\label{fig:RGB_A2744}
\end{figure*}

This Section presents another example of a gravitational arc simulated with our tool. 
In particular, we focus on the so-called System 3 \citep[][]{Vanzella_2022}, a source at $z = 3.98$ lensed by Abell 2744.
The mass model of this cluster is taken from \citet{bergamini23}.
The galaxy is assumed to have a total stellar mass $M_{*,tot} = 2 \times 10^8 M_{\odot}$, divided into a bulge with $M_{bul} = 5 \times 10^7 M_{\odot}$ and a disk with $M_{disk} = 1.5 \times 10^8 M_{\odot}$.
Here, we assume the $50\%$ of $M_{disk}$ goes into stellar clumps, for an overall mass of $7.5 \times 10^7 M_{\odot}$.
Moreover, to model the bulge and the disk as S\'ersic sources, we adopt the following parameters:

\begin{itemize}
    \item Bulge: S\'ersic index $n = 4$, effective radius $R_e = 0.1''$, position angle $\phi = \pi/8$;
    \item Disk: S\'ersic index $n = 1$, effective radius $R_e = 0.3''$, position angle $\phi = \pi/8$.
\end{itemize}
Both components are centered at source position $\beta_{s,1} = 28.395195 ''$ and $\beta_{s,2} = 40.2729467''$, and at redshift $z_{s} = 3.98$;
In order to build the spiral arms, we assume $R_h = 0.3''$, $A = 1.0$, $N_a = 4$, $\phi_d = 1.0$, $\alpha = 10.0$, $\phi = \frac{3}{4}\pi$. 

Regarding the stellar clumps, we use the same Schechter mass function ($\beta = -2.0$, $\delta = 1.0$, $\gamma = 1.0$, $x_{min} = 10^4 M_{\odot}$, $x_{max} = 10^{7} M_{\odot}$, and $x_{cut} = 5 \times 10^6 M_{\odot}$) and the same mass-size relation \citep[][]{brown2021} used in the previous example of the Cosmic Archipelago.
We also consider the same parameters, such as the S\'ersic index, position angle, and axis ratio, while the clumps' positions have been modified to suit the new host galaxy.

To use the Yggdrasil models, we consider the same input parameters ($Z = 0.004$, Kroupa IMF), SFH and ages for each component considered: the bulge has a SFR which is constant over 100 Myr and is assumed to be 1 Gyr old, the disk has the same SFR, but it is 500 Myr old, while for the clumps the SFR is constant over 10 Myr and their ages span between 1 Myr and 500 Myr.

Here we simulate observations with the Advanced Camera for Surveys (ACS) onboard HST (angular resolution: 0.05 arcsec/pix), and the usual HST WFC3/IR and JWST NIRCam.
Since HST WFC3/IR and JWST NIRCam have already been described in the previous example, we give some details only for HST ACS. 
The native pixel size of the ACS WFC camera onboard the HST is $ 0.0495''$/pix. 
We generate the PSF models using the public PSF modeling tool \texttt{Tiny Tim} \citep{tinytim}, simulating observations in the bands F435W, F606W, and F814W.

To define the sky background, we rely on the data of Abell 2744 from the HST proposals 11689 (PI Dupke) for ACS simulations and 13386 (PI Rodney) for WFC3/IR, while for JWST NIRCam, we consider the GTO program (PI Windhorst).
Following the method already described in the previous Section, we list all the $t_{exp}$ and $bkg$ values in Table~\ref{noise_Abell2744}.

\begin{table}[h!]
    \centering
    \captionsetup{position=above}
    \caption{As Table~\ref{noise_HST_WFC3}, but for HST ACS (F435W, F606W, and F814W), HST WFC3/IR (F105W, F125W, and F160W), and JWST NIRCam (F115W, F150W, F200W, F277W, F356W, and F444W).}
    \label{noise_Abell2744}
    \resizebox{\columnwidth}{!}{%
    \begin{tabular}{||c || c c c ||}
     \hline
         & F435W & F606W & F814W \\
        \hline\hline
        $t_{exp}$ [s] & 1284.0 & 1284.0 & 1284.0 \\ 
         \hline
         $bkg$ [counts/s/pix] & 0.0166096 & 0.0876545 & 0.0669051 \\
        \hline\hline
         & F105W & F125W & F160W \\
        \hline\hline
        $t_{exp}$ [s] & 652.94 & 602.93 & 702.94 \\ 
         \hline
         $bkg$ [counts/s/pix] & 0.297923 & 0.351991 & 0.271799 \\
        \hline\hline
         & F115W & F150W & F200W \\
        \hline\hline
        $t_{exp}$ [s] & 3349.872 & 2490.932 & 2490.932 \\ 
         \hline
         $bkg$ [counts/s/pix] & 0.228858 & 0.209639 & 0.17107 \\
         \hline\hline
         & F277W & F356W & F444W \\
        \hline\hline
        $t_{exp}$ [s] & 2490.932 & 2490.932 & 3349.872 \\ 
         \hline
         $bkg$ [counts/s/pix] & 0.220765 & 0.209639 & 0.228858 \\
         \hline
    \end{tabular}
    }
\end{table}

In Fig.~\ref{simul_A2744}, we show the observations simulated using the aforementioned combinations of filters and instruments. 
In this case we are able to detect many more clumps ($\rm >10$ in F115W) than in the previous example, despite the initial conditions of the two simulations being quite similar. We point out that in this case the simulated galaxy is made of two mirrored images, thus each of the individual clusters is imaged twice.
The observed difference with the Cosmic Archipelago is mainly driven by two factors: the difference in redshift and the one in magnification. The redshifts considered ($\rm z=3.980$ compared to 6.145 of the Cosmic Archipelago) correspond to a difference in observed magnitude of $\rm 1.1$ mag. The lensing magnification of the system in A2744 ranges from a factor $\sim30$ in the external regions (i.e. the left and right ends, in the lensed galaxy) to $>100$ near the critical line (toward the center of the lensed image, see \citealp{vanzella22}); on the other end, the largest magnifications achieved in the Cosmic Archipelago reach only a factor $\sim25$ \citep{Messa2025}. 
For instance, the clump highlighted in Fig. 7 is detectable in all the simulated observations, both with HST (with estimated S/N $\sim 11$ in F435W, $\sim 27$ in F606W, $\sim 15$ in F814W, $\sim 13$ in F105W, $\sim 4$ in F125W, and $\sim 8$ in F160W) and with JWST (reaching S/N $\sim 55$ in F115W, $\sim 45$ in F150W, $\sim 40$ in F200W, $\sim 22$ in F277W, $\sim 22$ in F356W, and $\sim 7$ in F444W). In contrast to the case discussed in Sect.~\ref{sec:disc}, here the significance never falls below the commonly adopted threshold of S/N = 3 (Messa et al. 2025), implying that the clump remains consistently detectable.

The top left panel of Fig.~\ref{fig:RGB_A2744} shows the high-resolution simulation of System 3 at $z = 3.98$ in Abell 2744. In this Figure the red, green and blue channels corresponds to rest-frame wavelengths $\lambda_{rf} = 1.0,~0.6$ and $0.4~\mu$m, respectively.
As done previously for Fig.~\ref{fig:RGB_M0416}, in addition to the lensed image at high-resolution, we provide the image of the convolution with the F115W PSF, without the inclusion of noise and background (top-right panel in Fig.~\ref{fig:RGB_A2744}). The middle row presents RGB images constructed from HST data: the left panel combines three ACS bands (red F814W, green F606W, and blue F435W), while the right one uses three WFC3/IR bands (red F160W, green F125W, and blue F105W). The bottom row shows RGB reconstructions from JWST NIRCam observations, using SW channels (red F200W, green F150W, and blue F115W) on the left and LW channels (red F444W, green F356W, and blue F277W) on the right.
As already pointed out in the simulations of the Cosmic Archipelago system in Section~\ref{sec:disc}, while the simulated high-resolution image clearly resolves numerous small-scale structures within the system, the lower resolution of the HST and JWST observations leads to significant blending, particularly among the faintest clumps. As a result, the observed emission is often dominated by the brightest components, potentially hiding a more complex distribution of stellar clusters. 

\section{Conclusions}
In this work, we have described the main phases and the fundamental ingredients for implementing a Python code to create realistic simulated observations of lensed, high-redshift, and star-forming galaxies. 
As examples, we present two simulations.
The first one, based on the same source size and lensing configuration of the real arc Cosmic Archipelago \citep[][]{Messa2025}, a source at $z = 6.145$ lensed by MACS J0416.1-2403, is used to describe the complete procedure of the tool. 
The second simulation follows a similar approach; however, in this case, we only present the final images and list the corresponding key parameters.
The pipeline can be summarized in the following fundamental steps:

\begin{itemize}
    \item Modeling of the host galaxy: the tool allows the user to choose which components to add to the source, including a disk, a bulge, and spiral arms. The brightness profiles of each component are described by S\'ersic brightness distributions. As input parameters, we have to define each component's stellar masses and sizes. Moreover, we have to define the source position on the source plane.
    \item Adding the stellar clumps: since we aim to simulate star-forming galaxies, our sources are designed to be populated by young stellar populations. The user selects which fraction of the host galaxy's stellar mass goes into stellar clumps. Then, the tool constructs them based on the following steps: it samples their masses following a Schechter distribution, places them according to the underlying host's brightness profile, and assigns them a radius that depends on their stellar mass.
    \item Simulating colors: the software can take as input parameters also the magnitudes in the bands that the user wishes to simulate. This step allows the simulation of different combinations of ages and SFHs for both the host galaxy and the stellar clumps. In the example described in this paper, we adopted a bulge of 1 Gyr and a disk of 500 Myr, while the clumps' ages span between 1 and 500 Myr. Regarding the SFH, we assume a SFR which is constant over 100 Myr for both components of the host galaxy, while for the clumps it is constant for 10 Myr. The tool relies on the Yggdrasil population synthesis code \citep[][]{Zackrisson_2011}.
    \item Simulating the lensing effects: by providing deflection angle maps, the pipeline can also include lensing effects. In the example presented here, we relied on the maps of MACS J0416.1-2403 from \texttt{Lenstool} \citep[][]{kneib96, jullo07, jullo09} by \citet{Bergamini_2023}.
    \item Including observational noises: observational noise and instrumental contamination can affect the previously generated noise-free images by providing the appropriate instrumental PSF, the flux from the sky background, and the proper exposure time, required to define the photon noise.
\end{itemize}

Following this framework, the tool can simulate realistic observations of lensed sources characterized by the presence of stellar clumps.
In Fig.~\ref{simul} and ~\ref{fig:RGB_M0416}, we present examples of multiband simulations for HST and JWST.
The reliability of our software has been tested and confirmed through a direct comparison with real images of the same sources, with a particular focus on the limiting magnitude, as discussed in Sect.~\ref{sec:disc}.
Sect. \ref{sec:a2744} describes and shows the simulations of the other example we considered in this work: the System 3 ($z = 3.98$) in A2744 \citep[][]{vanzella22}. 

The software presented in this work is a powerful and flexible tool designed to study the detectability of stellar clumps, particularly the smallest ones, in terms of mass and radius. It allows us to explore the fraction of clumps that appear blended into single and larger structures when, in reality, they consist of multiple smaller components. This approach enables us to probe the low-mass end of the stellar cluster IMF and to statistically investigate the physical properties of these systems.

One of the main strengths of the code lies in its modular design: all assumptions and parameters used to define the source galaxy, the gravitational lens configuration, the instrumental response, and the noise characteristics can be easily customized by the user. This makes the tool highly adaptable to a wide range of scientific goals and observational setups.

Beyond its immediate application in lensing simulations, the code can be used to generate realistic training datasets for machine learning algorithms aimed at detecting high-redshift clumps, or to benchmark and validate forward modeling frameworks such as GravityFM (Bergamini et al., in prep). Its flexibility and broad applicability make it a valuable resource for the community working on galaxy formation and gravitational lensing.

\section{Data availability}
The code used in this work is publicly available at \url{https://github.com/irenemini/clumPyLen-master}.

\begin{acknowledgements}
    The research activities described in this paper have been co-funded by the European Union – NextGenerationEU within PRIN 2022 project n.20229YBSAN - Globular clusters in cosmological simulations and in lensed fields: from their birth to the present epoch. MM and EV acknowledge support from PRIN-MUR 2020SKSTHZ. MM was supported by INAF Grant “The Big-Data era of cluster lensing”.
    LM acknowledges the financial contribution from the PRIN-MUR 2022 20227RNLY3 grant “The concordance cosmological model: stress-tests with galaxy clusters” supported by Next Generation EU and from the grant ASI n. 2024-10-HH.0 “Attività scientifiche per la missione Euclid – fase E”.
\end{acknowledgements}

\bibliographystyle{aa}
\bibliography{bibliography}

@ARTICLE{PortegiesZwart2010,
       author = {{Portegies Zwart}, Simon F. and {McMillan}, Stephen L.~W. and {Gieles}, Mark},
        title = "{Young Massive Star Clusters}",
      journal = {\araa},
         year = 2010,
        month = sep,
       volume = {48},
        pages = {431-493},
          doi = {10.1146/annurev-astro-081309-130834},
archivePrefix = {arXiv},
       eprint = {1002.1961},
 primaryClass = {astro-ph.GA},
       adsurl = {https://ui.adsabs.harvard.edu/abs/2010ARA&A..48..431P}
}

@ARTICLE{Zinnecker2007,
       author = {{Zinnecker}, Hans and {Yorke}, Harold W.},
        title = "{Toward Understanding Massive Star Formation}",
      journal = {\araa},
         year = 2007,
        month = sep,
       volume = {45},
       number = {1},
        pages = {481-563},
          doi = {10.1146/annurev.astro.44.051905.092549},
archivePrefix = {arXiv},
       eprint = {0707.1279},
 primaryClass = {astro-ph},
       adsurl = {https://ui.adsabs.harvard.edu/abs/2007ARA&A..45..481Z}
}

@ARTICLE{Krumholz2019,
       author = {{Krumholz}, Mark R. and {McKee}, Christopher F. and {Bland-Hawthorn}, Joss},
        title = "{Star Clusters Across Cosmic Time}",
      journal = {\araa},
         year = 2019,
        month = aug,
       volume = {57},
        pages = {227-303},
          doi = {10.1146/annurev-astro-091918-104430},
archivePrefix = {arXiv},
       eprint = {1812.01615},
 primaryClass = {astro-ph.GA},
       adsurl = {https://ui.adsabs.harvard.edu/abs/2019ARA&A..57..227K}
}

@ARTICLE{Adamo2023,
       author = {{Adamo}, Angela and {Usher}, Christopher and {Pfeffer}, Joel and {Claeyssens}, Ad{\'e}la{\"\i}de},
        title = "{The ages and metallicities of the globular clusters in the Sparkler}",
      journal = {\mnras},
         year = 2023,
        month = oct,
       volume = {525},
       number = {1},
        pages = {L6-L10},
          doi = {10.1093/mnrasl/slad084},
archivePrefix = {arXiv},
       eprint = {2306.11814},
 primaryClass = {astro-ph.GA},
       adsurl = {https://ui.adsabs.harvard.edu/abs/2023MNRAS.525L...6A}
}

@ARTICLE{Vanzella_2017,
       author = {{Vanzella}, E. and {Calura}, F. and {Meneghetti}, M. and {Mercurio}, A. and {Castellano}, M. and {Caminha}, G.~B. and {Balestra}, I. and {Rosati}, P. and {Tozzi}, P. and {De Barros}, S. and {Grazian}, A. and {D'Ercole}, A. and {Ciotti}, L. and {Caputi}, K. and {Grillo}, C. and {Merlin}, E. and {Pentericci}, L. and {Fontana}, A. and {Cristiani}, S. and {Coe}, D.},
        title = "{Paving the way for the JWST: witnessing globular cluster formation at z $>$ 3}",
      journal = {\mnras},
         year = 2017,
        month = jun,
       volume = {467},
       number = {4},
        pages = {4304-4321},
          doi = {10.1093/mnras/stx351},
archivePrefix = {arXiv},
       eprint = {1612.01526},
 primaryClass = {astro-ph.GA},
       adsurl = {https://ui.adsabs.harvard.edu/abs/2017MNRAS.467.4304V}
}

@ARTICLE{1976ApJ...203..297S,
       author = {{Schechter}, P.},
        title = "{An analytic expression for the luminosity function for galaxies.}",
      journal = {\apj},
         year = 1976,
        month = jan,
       volume = {203},
        pages = {297-306},
          doi = {10.1086/154079},
       adsurl = {https://ui.adsabs.harvard.edu/abs/1976ApJ...203..297S}
}

@ARTICLE{plezias18,
       author = {{Plazas}, A.~A. and {Meneghetti}, M. and {Maturi}, M. and {Rhodes}, J.},
        title = "{Image simulations for gravitational lensing with SKYLENS}",
      journal = {\mnras},
         year = 2019,
        month = jan,
       volume = {482},
       number = {2},
        pages = {2823-2832},
          doi = {10.1093/mnras/sty2737},
archivePrefix = {arXiv},
       eprint = {1805.05481},
 primaryClass = {astro-ph.CO},
       adsurl = {https://ui.adsabs.harvard.edu/abs/2019MNRAS.482.2823P}
}

@ARTICLE{Adamo2020,
       author = {{Adamo}, A. and {Hollyhead}, K. and {Messa}, M. and {Ryon}, J.~E. and {Bajaj}, V. and {Runnholm}, A. and {Aalto}, S. and {Calzetti}, D. and {Gallagher}, J.~S. and {Hayes}, M.~J. and {Kruijssen}, J.~M.~D. and {K{\"o}nig}, S. and {Larsen}, S.~S. and {Melinder}, J. and {Sabbi}, E. and {Smith}, L.~J. and {{\"O}stlin}, G.},
        title = "{Star cluster formation in the most extreme environments: insights from the HiPEEC survey}",
      journal = {\mnras},
         year = 2020,
        month = dec,
       volume = {499},
       number = {3},
        pages = {3267-3294},
          doi = {10.1093/mnras/staa2380},
archivePrefix = {arXiv},
       eprint = {2008.12794},
 primaryClass = {astro-ph.GA},
       adsurl = {https://ui.adsabs.harvard.edu/abs/2020MNRAS.499.3267A}
}

@ARTICLE{Ryon2015,
       author = {{Ryon}, J.~E. and {Bastian}, N. and {Adamo}, A. and {Konstantopoulos}, I.~S. and {Gallagher}, J.~S. and {Larsen}, S. and {Hollyhead}, K. and {Silva-Villa}, E. and {Smith}, L.~J.},
        title = "{Sizes and shapes of young star cluster light profiles in M83}",
      journal = {\mnras},
         year = 2015,
        month = sep,
       volume = {452},
       number = {1},
        pages = {525-539},
          doi = {10.1093/mnras/stv1282},
archivePrefix = {arXiv},
       eprint = {1506.02042},
 primaryClass = {astro-ph.GA},
       adsurl = {https://ui.adsabs.harvard.edu/abs/2015MNRAS.452..525R}
}

@ARTICLE{ryon2017,
       author = {{Ryon}, J.~E. and {Gallagher}, J.~S. and {Smith}, L.~J. and {Adamo}, A. and {Calzetti}, D. and {Bright}, S.~N. and {Cignoni}, M. and {Cook}, D.~O. and {Dale}, D.~A. and {Elmegreen}, B.~E. and {Fumagalli}, M. and {Gouliermis}, D.~A. and {Grasha}, K. and {Grebel}, E.~K. and {Kim}, H. and {Messa}, M. and {Thilker}, D. and {Ubeda}, L.},
        title = "{Effective Radii of Young, Massive Star Clusters in Two LEGUS Galaxies}",
      journal = {\apj},
         year = 2017,
        month = jun,
       volume = {841},
       number = {2},
          eid = {92},
        pages = {92},
          doi = {10.3847/1538-4357/aa719e},
archivePrefix = {arXiv},
       eprint = {1705.02692},
 primaryClass = {astro-ph.GA},
       adsurl = {https://ui.adsabs.harvard.edu/abs/2017ApJ...841...92R}
}

@ARTICLE{brown2021,
       author = {{Brown}, Gillen and {Gnedin}, Oleg Y.},
        title = "{Radii of young star clusters in nearby galaxies}",
      journal = {\mnras},
         year = 2021,
        month = dec,
       volume = {508},
       number = {4},
        pages = {5935-5953},
          doi = {10.1093/mnras/stab2907},
archivePrefix = {arXiv},
       eprint = {2106.12420},
 primaryClass = {astro-ph.GA},
       adsurl = {https://ui.adsabs.harvard.edu/abs/2021MNRAS.508.5935B}
}

@ARTICLE{kneib96,
       author = {{Kneib}, J. -P. and {Ellis}, R.~S. and {Smail}, I. and {Couch}, W.~J. and {Sharples}, R.~M.},
        title = "{Hubble Space Telescope Observations of the Lensing Cluster Abell 2218}",
      journal = {\apj},
         year = 1996,
        month = nov,
       volume = {471},
        pages = {643},
          doi = {10.1086/177995},
archivePrefix = {arXiv},
       eprint = {astro-ph/9511015},
 primaryClass = {astro-ph},
       adsurl = {https://ui.adsabs.harvard.edu/abs/1996ApJ...471..643K}
}

@ARTICLE{jullo07,
       author = {{Jullo}, E. and {Kneib}, J. -P. and {Limousin}, M. and {El{\'\i}asd{\'o}ttir}, {\'A}. and {Marshall}, P.~J. and {Verdugo}, T.},
        title = "{A Bayesian approach to strong lensing modelling of galaxy clusters}",
      journal = {New Journal of Physics},
         year = 2007,
        month = dec,
       volume = {9},
       number = {12},
        pages = {447},
          doi = {10.1088/1367-2630/9/12/447},
archivePrefix = {arXiv},
       eprint = {0706.0048},
 primaryClass = {astro-ph},
       adsurl = {https://ui.adsabs.harvard.edu/abs/2007NJPh....9..447J}
}

@ARTICLE{jullo09,
       author = {{Jullo}, E. and {Kneib}, J. -P.},
        title = "{Multiscale cluster lens mass mapping - I. Strong lensing modelling}",
      journal = {\mnras},
         year = 2009,
        month = may,
       volume = {395},
       number = {3},
        pages = {1319-1332},
          doi = {10.1111/j.1365-2966.2009.14654.x},
archivePrefix = {arXiv},
       eprint = {0901.3792},
 primaryClass = {astro-ph.CO},
       adsurl = {https://ui.adsabs.harvard.edu/abs/2009MNRAS.395.1319J}
}

@ARTICLE{bergamini23,
       author = {{Bergamini}, P. and {Acebron}, A. and {Grillo}, C. and {Rosati}, P. and {Caminha}, G.~B. and {Mercurio}, A. and {Vanzella}, E. and {Angora}, G. and {Brammer}, G. and {Meneghetti}, M. and {Nonino}, M.},
        title = "{New high-precision strong lensing modeling of Abell 2744. Preparing for JWST observations}",
      journal = {\aap},
         year = 2023,
        month = feb,
       volume = {670},
          eid = {A60},
        pages = {A60},
          doi = {10.1051/0004-6361/202244575},
archivePrefix = {arXiv},
       eprint = {2207.09416},
 primaryClass = {astro-ph.CO},
       adsurl = {https://ui.adsabs.harvard.edu/abs/2023A&A...670A..60B}
}

@article{Bergamini_2023,
   title={A state-of-the-art strong-lensing model of MACS J0416.1−2403 with the largest sample of spectroscopic multiple images},
   volume={674},
   ISSN={1432-0746},
   url={http://dx.doi.org/10.1051/0004-6361/202244834},
   DOI={10.1051/0004-6361/202244834},
   journal={\aap},
   publisher={EDP Sciences},
   author={Bergamini, P. and Grillo, C. and Rosati, P. and Vanzella, E. and Meštrić, U. and Mercurio, A. and Acebron, A. and Caminha, G. B. and Granata, G. and Meneghetti, M. and Angora, G. and Nonino, M.},
   year={2023},
   month=jun, pages={A79} 
}

@ARTICLE{bergamini19,
       author = {{Bergamini}, P. and {Rosati}, P. and {Mercurio}, A. and {Grillo}, C. and {Caminha}, G.~B. and {Meneghetti}, M. and {Agnello}, A. and {Biviano}, A. and {Calura}, F. and {Giocoli}, C. and {Lombardi}, M. and {Rodighiero}, G. and {Vanzella}, E.},
        title = "{Enhanced cluster lensing models with measured galaxy kinematics}",
      journal = {\aap},
         year = 2019,
        month = nov,
       volume = {631},
          eid = {A130},
        pages = {A130},
          doi = {10.1051/0004-6361/201935974},
archivePrefix = {arXiv},
       eprint = {1905.13236},
 primaryClass = {astro-ph.GA},
       adsurl = {https://ui.adsabs.harvard.edu/abs/2019A&A...631A.130B}
}

@ARTICLE{bergamini20,
       author = {{Bergamini}, P. and {Rosati}, P. and {Vanzella}, E. and {Caminha}, G.~B. and {Grillo}, C. and {Mercurio}, A. and {Meneghetti}, M. and {Angora}, G. and {Calura}, F. and {Nonino}, M. and {Tozzi}, P.},
        title = "{A new high-precision strong lensing model of the galaxy cluster MACS J0416.1-2403. Robust characterization of the cluster mass distribution from VLT/MUSE deep observations}",
      journal = {\aap},
         year = 2021,
        month = jan,
       volume = {645},
          eid = {A140},
        pages = {A140},
          doi = {10.1051/0004-6361/202039564},
archivePrefix = {arXiv},
       eprint = {2010.00027},
 primaryClass = {astro-ph.GA},
       adsurl = {https://ui.adsabs.harvard.edu/abs/2021A&A...645A.140B}
}

@ARTICLE{Meneghetti2008,
       author = {{Meneghetti}, M. and {Melchior}, P. and {Grazian}, A. and {De Lucia}, G. and {Dolag}, K. and {Bartelmann}, M. and {Heymans}, C. and {Moscardini}, L. and {Radovich}, M.},
        title = "{Realistic simulations of gravitational lensing by galaxy clusters: extracting arc parameters from mock DUNE images}",
      journal = {\aap},
         year = 2008,
        month = may,
       volume = {482},
       number = {2},
        pages = {403-418},
          doi = {10.1051/0004-6361:20079119},
archivePrefix = {arXiv},
       eprint = {0711.3418},
 primaryClass = {astro-ph},
       adsurl = {https://ui.adsabs.harvard.edu/abs/2008A&A...482..403M}
}

@ARTICLE{cava2018,
       author = {{Cava}, Antonio and {Schaerer}, Daniel and {Richard}, Johan and {P{\'e}rez-Gonz{\'a}lez}, Pablo G. and {Dessauges-Zavadsky}, Miroslava and {Mayer}, Lucio and {Tamburello}, Valentina},
        title = "{The nature of giant clumps in distant galaxies probed by the anatomy of the cosmic snake}",
      journal = {Nature Astronomy},
         year = 2018,
        month = nov,
       volume = {2},
        pages = {76-82},
          doi = {10.1038/s41550-017-0295-x},
archivePrefix = {arXiv},
       eprint = {1711.03977},
 primaryClass = {astro-ph.GA},
       adsurl = {https://ui.adsabs.harvard.edu/abs/2018NatAs...2...76C}
}

@ARTICLE{vanzella22,
       author = {{Vanzella}, E. and {Castellano}, M. and {Bergamini}, P. and {Treu}, T. and {Mercurio}, A. and {Scarlata}, C. and {Rosati}, P. and {Grillo}, C. and {Acebron}, A. and {Caminha}, G.~B. and {Nonino}, M. and {Nanayakkara}, T. and {Roberts-Borsani}, G. and {Bradac}, M. and {Wang}, X. and {Brammer}, G. and {Strait}, V. and {Vulcani}, B. and {Me{\v{s}}tri{\'c}}, U. and {Meneghetti}, M. and {Calura}, F. and {Henry}, Alaina and {Zanella}, A. and {Trenti}, M. and {Boyett}, K. and {Morishita}, T. and {Calabr{\`o}}, A. and {Glazebrook}, K. and {Marchesini}, D. and {Birrer}, S. and {Yang}, L. and {Jones}, T.},
        title = "{Early Results from GLASS-JWST. VII. Evidence for Lensed, Gravitationally Bound Protoglobular Clusters at z = 4 in the Hubble Frontier Field A2744}",
      journal = {\apjl},
         year = 2022,
        month = dec,
       volume = {940},
       number = {2},
          eid = {L53},
        pages = {L53},
          doi = {10.3847/2041-8213/ac8c2d},
archivePrefix = {arXiv},
       eprint = {2208.00520},
 primaryClass = {astro-ph.GA},
       adsurl = {https://ui.adsabs.harvard.edu/abs/2022ApJ...940L..53V}
}

@INPROCEEDINGS{tinytim,
       author = {{Krist}, John E. and {Hook}, Richard N. and {Stoehr}, Felix},
        title = "{20 years of Hubble Space Telescope optical modeling using Tiny Tim}",
    booktitle = {Optical Modeling and Performance Predictions V},
         year = 2011,
       editor = {{Kahan}, Mark A.},
       series = {Society of Photo-Optical Instrumentation Engineers (SPIE) Conference Series},
       volume = {8127},
        month = oct,
          eid = {81270J},
        pages = {81270J},
          doi = {10.1117/12.892762},
       adsurl = {https://ui.adsabs.harvard.edu/abs/2011SPIE.8127E..0JK}
}

@article{Adamo_2020,
	doi = {10.1007/s11214-020-00690-x},
  
	url = {https://doi.org/10.1007%2Fs11214-020-00690-x},
  
	year = 2020,
	month = {jun},
  
	publisher = {Springer Science and Business Media {LLC}
},
  
	volume = {216},
  
	number = {4},
  
	author = {Angela Adamo and Peter Zeidler and J. M. Diederik Kruijssen and M{\'{e}}lanie Chevance and Mark Gieles and Daniela Calzetti and Corinne Charbonnel and Hans Zinnecker and Martin G. H. Krause},
  
	title = {Star Clusters Near and Far},
  
	journal = {Space Science Reviews}
}

@article{Vanzella_2019,
	doi = {10.1093/mnras/stz2286},
  
	url = {https://doi.org/10.1093%2Fmnras%2Fstz2286},
  
	year = 2019,
	month = {aug},
  
	publisher = {Oxford University Press ({OUP})},
  
	volume = {491},
  
	number = {1},
  
	pages = {1093--1103},
  
	author = {E Vanzella and G B Caminha and F Calura and G Cupani and M Meneghetti and M Castellano and P Rosati and A Mercurio and E Sani and C Grillo and R Gilli and M Mignoli and A Comastri and M Nonino and S Cristiani and M Giavalisco and K Caputi},
  
	title = {Ionizing the intergalactic medium by star clusters: the first empirical evidence},
  
	journal = {Monthly Notices of the Royal Astronomical Society}
}

@article{Vanzella_2023,
	doi = {10.3847/1538-4357/acb59a},
	url = {https://doi.org/10.3847%2F1538-4357%2Facb59a},
	year = 2023,
	month = {mar},
	publisher = {American Astronomical Society},
	volume = {945},
	number = {1},
	pages = {53},
	author = {Eros Vanzella and Ad{\'{e}
}laïde Claeyssens and Brian Welch and Angela Adamo and Dan Coe and Jose M. Diego and Guillaume Mahler and Gourav Khullar and Vasily Kokorev and Masamune Oguri and Swara Ravindranath and Lukas J. Furtak and Tiger Yu-Yang Hsiao and  Abdurro'uf and Nir Mandelker and Gabriel Brammer and Larry D. Bradley and Maru{\v{s}}a Brada{\v{c}} and Christopher J. Conselice and Pratika Dayal and Mario Nonino and Felipe Andrade-Santos and Rogier A. Windhorst and Nor Pirzkal and Keren Sharon and S. E. de Mink and Seiji Fujimoto and Adi Zitrin and Jan J. Eldridge and Colin Norman},
	title = {{JWST}/{NIRCam} Probes Young Star Clusters in the Reionization Era Sunrise Arc},
	journal = {The Astrophysical Journal}
}

@article{Vanzella_2022,
	doi = {10.1051/0004-6361/202141590},
  
	url = {https://doi.org/10.1051%2F0004-6361%2F202141590},
  
	year = 2022,
	month = {feb},
  
	publisher = {{EDP} Sciences},
  
	volume = {659},
  
	pages = {A2},
  
	author = {E. Vanzella and M. Castellano and P. Bergamini and M. Meneghetti and A. Zanella and F. Calura and G. B. Caminha and P. Rosati and G. Cupani and U. Me{\v{s}
}tri{\'{c}} and G. Brammer and P. Tozzi and A. Mercurio and C. Grillo and E. Sani and S. Cristiani and M. Nonino and E. Merlin and G. V. Pignataro},
  
	title = {High star cluster formation efficiency in the strongly lensed Sunburst Lyman-continuum galaxy at $\less$i$\greater$z$\less$/i$\greater$ = 2.37},
  
	journal = {\aap}
}

@BOOK{2021LNP...956.....M,
       author = {{Meneghetti}, Massimo},
        title = "{Introduction to Gravitational Lensing; With Python Examples}",
         year = 2021,
        publisher = {Springer}, 
       volume = {956},
          doi = {10.1007/978-3-030-73582-1},
       adsurl = {https://ui.adsabs.harvard.edu/abs/2021LNP...956.....M}
}

@article{Metcalf_2019,
   title={The strong gravitational lens finding challenge},
   volume={625},
   ISSN={1432-0746},
   url={http://dx.doi.org/10.1051/0004-6361/201832797},
   DOI={10.1051/0004-6361/201832797},
   journal={\aap},
   publisher={EDP Sciences},
   author={Metcalf, R. B. and Meneghetti, M. and Avestruz, C. and Bellagamba, F. and Bom, C. R. and Bertin, E. and Cabanac, R. and Courbin, F. and Davies, A. and Decencière, E. and Flamary, R. and Gavazzi, R. and Geiger, M. and Hartley, P. and Huertas-Company, M. and Jackson, N. and Jacobs, C. and Jullo, E. and Kneib, J.-P. and Koopmans, L. V. E. and Lanusse, F. and Li, C.-L. and Ma, Q. and Makler, M. and Li, N. and Lightman, M. and Petrillo, C. E. and Serjeant, S. and Schäfer, C. and Sonnenfeld, A. and Tagore, A. and Tortora, C. and Tuccillo, D. and Valentín, M. B. and Velasco-Forero, S. and Verdoes Kleijn, G. A. and Vernardos, G.},
   year={2019},
   month=may, pages={A119} }

@article{Zackrisson_2011,
   title={THE SPECTRAL EVOLUTION OF THE FIRST GALAXIES. I.JAMES WEBB SPACE TELESCOPEDETECTION LIMITS AND COLOR CRITERIA FOR POPULATION III GALAXIES},
   volume={740},
   ISSN={1538-4357},
   url={http://dx.doi.org/10.1088/0004-637X/740/1/13},
   DOI={10.1088/0004-637x/740/1/13},
   number={1},
   journal={The Astrophysical Journal},
   publisher={American Astronomical Society},
   author={Zackrisson, Erik and Rydberg, Claes-Erik and Schaerer, Daniel and Östlin, Göran and Tuli, Manan},
   year={2011},
   month=sep, pages={13} }

@ARTICLE{2001MNRAS.322..231K,
       author = {{Kroupa}, Pavel},
        title = "{On the variation of the initial mass function}",
      journal = {\mnras},
         year = 2001,
        month = apr,
       volume = {322},
       number = {2},
        pages = {231-246},
          doi = {10.1046/j.1365-8711.2001.04022.x},
archivePrefix = {arXiv},
       eprint = {astro-ph/0009005},
 primaryClass = {astro-ph},
       adsurl = {https://ui.adsabs.harvard.edu/abs/2001MNRAS.322..231K}
}

@ARTICLE{1963BAAA....6...41S,
       author = {{S{\'e}rsic}, J.~L.},
        title = "{Influence of the atmospheric and instrumental dispersion on the brightness distribution in a galaxy}",
      journal = {Boletin de la Asociacion Argentina de Astronomia La Plata Argentina},
         year = 1963,
        month = feb,
       volume = {6},
        pages = {41-43},
       adsurl = {https://ui.adsabs.harvard.edu/abs/1963BAAA....6...41S}
}

@ARTICLE{2010RSPTA.368..867L,
       author = {{Larsen}, S.~S.},
        title = "{Young and intermediate-age massive star clusters}",
      journal = {Philosophical Transactions of the Royal Society of London Series A},
         year = 2010,
        month = jan,
       volume = {368},
       number = {1913},
        pages = {867-887},
          doi = {10.1098/rsta.2009.0255},
archivePrefix = {arXiv},
       eprint = {0911.0796},
 primaryClass = {astro-ph.GA},
       adsurl = {https://ui.adsabs.harvard.edu/abs/2010RSPTA.368..867L}
}

@ARTICLE{Mowla_2022,
       author = {{Mowla}, Lamiya and {Iyer}, Kartheik G. and {Desprez}, Guillaume and {Estrada-Carpenter}, Vicente and {Martis}, Nicholas S. and {Noirot}, Ga{\"e}l and {Sarrouh}, Ghassan T. and {Strait}, Victoria and {Asada}, Yoshihisa and {Abraham}, Roberto G. and {Brammer}, Gabriel and {Sawicki}, Marcin and {Willott}, Chris J. and {Bradac}, Marusa and {Doyon}, Ren{\'e} and {Muzzin}, Adam and {Pacifici}, Camilla and {Ravindranath}, Swara and {Zabl}, Johannes},
        title = "{The Sparkler: Evolved High-redshift Globular Cluster Candidates Captured by JWST}",
      journal = {\apjl},
         year = 2022,
        month = oct,
       volume = {937},
       number = {2},
          eid = {L35},
        pages = {L35},
          doi = {10.3847/2041-8213/ac90ca},
archivePrefix = {arXiv},
       eprint = {2208.02233},
 primaryClass = {astro-ph.GA},
       adsurl = {https://ui.adsabs.harvard.edu/abs/2022ApJ...937L..35M}
}

@ARTICLE{Claeyssens2023,
       author = {{Claeyssens}, Ad{\'e}la{\"\i}de and {Adamo}, Angela and {Richard}, Johan and {Mahler}, Guillaume and {Messa}, Matteo and {Dessauges-Zavadsky}, Miroslava},
        title = "{Star formation at the smallest scales: a JWST study of the clump populations in SMACS0723}",
      journal = {\mnras},
         year = 2023,
        month = apr,
       volume = {520},
       number = {2},
        pages = {2180-2203},
          doi = {10.1093/mnras/stac3791},
archivePrefix = {arXiv},
       eprint = {2208.10450},
 primaryClass = {astro-ph.GA},
       adsurl = {https://ui.adsabs.harvard.edu/abs/2023MNRAS.520.2180C}
}

@ARTICLE{mowla2024fireflysparkleearlieststages,
       author = {{Mowla}, Lamiya and {Iyer}, Kartheik and {Asada}, Yoshihisa and {Desprez}, Guillaume and {Tan}, Vivian Yun Yan and {Martis}, Nicholas and {Sarrouh}, Ghassan and {Strait}, Victoria and {Abraham}, Roberto and {Brada{\v{c}}}, Maru{\v{s}}a and {Brammer}, Gabriel and {Muzzin}, Adam and {Pacifici}, Camilla and {Ravindranath}, Swara and {Sawicki}, Marcin and {Willott}, Chris and {Estrada-Carpenter}, Vince and {Jahan}, Nusrath and {Noirot}, Ga{\"e}l and {Matharu}, Jasleen and {Rihtar{\v{s}}i{\v{c}}}, Gregor and {Zabl}, Johannes},
        title = "{Formation of a low-mass galaxy from star clusters in a 600-million-year-old Universe}",
      journal = {\nat},
         year = 2024,
        month = dec,
       volume = {636},
       number = {8042},
        pages = {332-336},
          doi = {10.1038/s41586-024-08293-0},
archivePrefix = {arXiv},
       eprint = {2402.08696},
 primaryClass = {astro-ph.GA},
       adsurl = {https://ui.adsabs.harvard.edu/abs/2024Natur.636..332M}
}

@ARTICLE{adamo2024boundstarclustersobserved,
       author = {{Adamo}, Angela and {Bradley}, Larry D. and {Vanzella}, Eros and {Claeyssens}, Ad{\'e}la{\"\i}de and {Welch}, Brian and {Diego}, Jose M. and {Mahler}, Guillaume and {Oguri}, Masamune and {Sharon}, Keren and {Abdurro'uf} and {Hsiao}, Tiger Yu-Yang and {Xu}, Xinfeng and {Messa}, Matteo and {Lassen}, Augusto E. and {Zackrisson}, Erik and {Brammer}, Gabriel and {Coe}, Dan and {Kokorev}, Vasily and {Ricotti}, Massimo and {Zitrin}, Adi and {Fujimoto}, Seiji and {Inoue}, Akio K. and {Resseguier}, Tom and {Rigby}, Jane R. and {Jim{\'e}nez-Teja}, Yolanda and {Windhorst}, Rogier A. and {Hashimoto}, Takuya and {Tamura}, Yoichi},
        title = "{Bound star clusters observed in a lensed galaxy 460 Myr after the Big Bang}",
      journal = {\nat},
         year = 2024,
        month = aug,
       volume = {632},
       number = {8025},
        pages = {513-516},
          doi = {10.1038/s41586-024-07703-7},
archivePrefix = {arXiv},
       eprint = {2401.03224},
 primaryClass = {astro-ph.GA},
       adsurl = {https://ui.adsabs.harvard.edu/abs/2024Natur.632..513A}
}

@ARTICLE{2018ApJ...864L..22S,
       author = {{Salmon}, Brett and {Coe}, Dan and {Bradley}, Larry and {Brada{\v{c}}}, Marusa and {Strait}, Victoria and {Paterno-Mahler}, Rachel and {Huang}, Kuang-Han and {Oesch}, Pascal A. and {Zitrin}, Adi and {Acebron}, Ana and {Cibirka}, Nath{\'a}lia and {Kikuchihara}, Shotaro and {Oguri}, Masamune and {Brammer}, Gabriel B. and {Sharon}, Keren and {Trenti}, Michele and {Avila}, Roberto J. and {Ogaz}, Sara and {Andrade-Santos}, Felipe and {Carrasco}, Daniela and {Cerny}, Catherine and {Dawson}, William and {Frye}, Brenda L. and {Hoag}, Austin and {Jones}, Christine and {Mainali}, Ramesh and {Ouchi}, Masami and {Rodney}, Steven A. and {Stark}, Daniel and {Umetsu}, Keiichi},
        title = "{RELICS: A Candidate z {\ensuremath{\sim}} 10 Galaxy Strongly Lensed into a Spatially Resolved Arc}",
      journal = {\apjl},
         year = 2018,
        month = sep,
       volume = {864},
       number = {1},
          eid = {L22},
        pages = {L22},
          doi = {10.3847/2041-8213/aadc10},
archivePrefix = {arXiv},
       eprint = {1801.03103},
 primaryClass = {astro-ph.GA},
       adsurl = {https://ui.adsabs.harvard.edu/abs/2018ApJ...864L..22S}
}

@ARTICLE{elmegreen2006,
       author = {{Elmegreen}, Bruce G. and {Elmegreen}, Debra Meloy},
        title = "{Observations of Thick Disks in the Hubble Space Telescope Ultra Deep Field}",
      journal = {\apj},
         year = 2006,
        month = oct,
       volume = {650},
       number = {2},
        pages = {644-660},
          doi = {10.1086/507578},
archivePrefix = {arXiv},
       eprint = {astro-ph/0607540},
 primaryClass = {astro-ph},
       adsurl = {https://ui.adsabs.harvard.edu/abs/2006ApJ...650..644E}
}

@article{wisnioski2012,
    author = {Wisnioski, Emily and Glazebrook, Karl and Blake, Chris and Poole, Gregory B. and Green, Andrew W. and Wyder, Ted and Martin, Chris},
    title = "{Scaling relations of star-forming regions: from kpc-sized clumps to H ii regions}",
    journal = {Monthly Notices of the Royal Astronomical Society},
    volume = {422},
    number = {4},
    pages = {3339-3355},
    year = {2012},
    month = {05},
    issn = {0035-8711},
    doi = {10.1111/j.1365-2966.2012.20850.x},
    url = {https://doi.org/10.1111/j.1365-2966.2012.20850.x},
    eprint = {https://academic.oup.com/mnras/article-pdf/422/4/3339/18604040/mnras0422-3339.pdf},
}

@ARTICLE{johnson2017,
       author = {{Johnson}, Traci L. and {Rigby}, Jane R. and {Sharon}, Keren and {Gladders}, Michael D. and {Florian}, Michael and {Bayliss}, Matthew B. and {Wuyts}, Eva and {Whitaker}, Katherine E. and {Livermore}, Rachael and {Murray}, Katherine T.},
        title = "{Star Formation at z = 2.481 in the Lensed Galaxy SDSS J1110+6459: Star Formation Down to 30 pc Scales}",
      journal = {\apjl},
         year = 2017,
        month = jul,
       volume = {843},
       number = {2},
          eid = {L21},
        pages = {L21},
          doi = {10.3847/2041-8213/aa7516},
archivePrefix = {arXiv},
       eprint = {1707.00706},
 primaryClass = {astro-ph.GA},
       adsurl = {https://ui.adsabs.harvard.edu/abs/2017ApJ...843L..21J}
}

@ARTICLE{Mestric2022,
       author = {{Me{\v{s}}tri{\'c}}, U. and {Vanzella}, E. and {Zanella}, A. and {Castellano}, M. and {Calura}, F. and {Rosati}, P. and {Bergamini}, P. and {Mercurio}, A. and {Meneghetti}, M. and {Grillo}, C. and {Caminha}, G.~B. and {Nonino}, M. and {Merlin}, E. and {Cupani}, G. and {Sani}, E.},
        title = "{Exploring the physical properties of lensed star-forming clumps at 2 {\ensuremath{\lesssim}} z {\ensuremath{\lesssim}} 6}",
      journal = {\mnras},
         year = 2022,
        month = nov,
       volume = {516},
       number = {3},
        pages = {3532-3555},
          doi = {10.1093/mnras/stac2309},
archivePrefix = {arXiv},
       eprint = {2202.09377},
 primaryClass = {astro-ph.GA},
       adsurl = {https://ui.adsabs.harvard.edu/abs/2022MNRAS.516.3532M}
}

@ARTICLE{Messa2022,
       author = {{Messa}, Matteo and {Dessauges-Zavadsky}, Miroslava and {Richard}, Johan and {Adamo}, Angela and {Nagy}, David and {Combes}, Fran{\c{c}}oise and {Mayer}, Lucio and {Ebeling}, Harald},
        title = "{Multiply lensed star forming clumps in the A521-sys1 galaxy at redshift 1}",
      journal = {\mnras},
         year = 2022,
        month = oct,
       volume = {516},
       number = {2},
        pages = {2420-2443},
          doi = {10.1093/mnras/stac2189},
archivePrefix = {arXiv},
       eprint = {2208.02863},
 primaryClass = {astro-ph.GA},
       adsurl = {https://ui.adsabs.harvard.edu/abs/2022MNRAS.516.2420M}
}

@ARTICLE{Messa2024a,
       author = {{Messa}, Matteo and {Dessauges-Zavadsky}, Miroslava and {Adamo}, Angela and {Richard}, Johan and {Claeyssens}, Ad{\'e}la{\"\i}de},
        title = "{Properties of the brightest young stellar clumps in extremely lensed galaxies at redshifts 4 to 5}",
      journal = {\mnras},
         year = 2024,
        month = apr,
       volume = {529},
       number = {3},
        pages = {2162-2179},
          doi = {10.1093/mnras/stae565},
archivePrefix = {arXiv},
       eprint = {2402.14920},
 primaryClass = {astro-ph.GA},
       adsurl = {https://ui.adsabs.harvard.edu/abs/2024MNRAS.529.2162M}
}

@ARTICLE{Messa2025,
       author = {{Messa}, M. and {Vanzella}, E. and {Loiacono}, F. and {Bergamini}, P. and {Castellano}, M. and {Sun}, B. and {Willott}, C. and {Windhorst}, R.~A. and {Yan}, H. and {Angora}, G. and {Rosati}, P. and {Adamo}, A. and {Annibali}, F. and {Bolamperti}, A. and {Brada{\v{c}}}, M. and {Bradley}, L.~D. and {Calura}, F. and {Claeyssens}, A. and {Comastri}, A. and {Conselice}, C.~J. and {D'Silva}, J.~C.~J. and {Dickinson}, M. and {Frye}, B.~L. and {Grillo}, C. and {Grogin}, N.~A. and {Gruppioni}, C. and {Koekemoer}, A.~M. and {Meneghetti}, M. and {Me{\v{s}}tri{\'c}}, U. and {Pascale}, R. and {Ravindranath}, S. and {Ricotti}, M. and {Summers}, J. and {Zanella}, A.},
        title = "{Anatomy of a z = 6 Lyman-{\ensuremath{\alpha}} emitter down to parsec scales: Extreme UV slopes, metal-poor regions, and possibly leaking star clusters}",
      journal = {\aap},
         year = 2025,
        month = feb,
       volume = {694},
          eid = {A59},
        pages = {A59},
          doi = {10.1051/0004-6361/202451695},
archivePrefix = {arXiv},
       eprint = {2407.20331},
 primaryClass = {astro-ph.GA},
       adsurl = {https://ui.adsabs.harvard.edu/abs/2025A&A...694A..59M}
}

@ARTICLE{Ji2024,
       author = {{Ji}, Zhiyuan and {Williams}, Christina C. and {Tacchella}, Sandro and {Suess}, Katherine A. and {Baker}, William M. and {Alberts}, Stacey and {Bunker}, Andrew J. and {Johnson}, Benjamin D. and {Robertson}, Brant and {Sun}, Fengwu and {Eisenstein}, Daniel J. and {Rieke}, Marcia and {Maseda}, Michael V. and {Hainline}, Kevin and {Hausen}, Ryan and {Rieke}, George and {Willmer}, Christopher N.~A. and {Egami}, Eiichi and {Shivaei}, Irene and {Carniani}, Stefano and {Charlot}, Stephane and {Chevallard}, Jacopo and {Curtis-Lake}, Emma and {Looser}, Tobias J. and {Maiolino}, Roberto and {Willott}, Chris and {Chen}, Zuyi and {Helton}, Jakob M. and {Lyu}, Jianwei and {Nelson}, Erica and {Bhatawdekar}, Rachana and {Boyett}, Kristan and {Sandles}, Lester},
        title = "{JADES + JEMS: A Detailed Look at the Buildup of Central Stellar Cores and Suppression of Star Formation in Galaxies at Redshifts 3 < z < 4.5}",
      journal = {\apj},
         year = 2024,
        month = oct,
       volume = {974},
       number = {1},
          eid = {135},
        pages = {135},
          doi = {10.3847/1538-4357/ad6e7f},
archivePrefix = {arXiv},
       eprint = {2305.18518},
 primaryClass = {astro-ph.GA},
       adsurl = {https://ui.adsabs.harvard.edu/abs/2024ApJ...974..135J}
}

@ARTICLE{kalita2025b,
       author = {{Kalita}, Boris S. and {Silverman}, John D. and {Daddi}, Emanuele and {Mercier}, Wilfried and {Ho}, Luis C. and {Ding}, Xuheng},
        title = "{Near-IR clumps and their properties in high-z galaxies with JWST/NIRCam}",
      journal = {\mnras},
         year = 2025,
        month = feb,
       volume = {537},
       number = {1},
        pages = {402-418},
          doi = {10.1093/mnras/staf031},
archivePrefix = {arXiv},
       eprint = {2402.02679},
 primaryClass = {astro-ph.GA},
       adsurl = {https://ui.adsabs.harvard.edu/abs/2025MNRAS.537..402K}
}

@ARTICLE{kalita2025a,
       author = {{Kalita}, Boris S. and {Suzuki}, Tomoko L. and {Kashino}, Daichi and {Silverman}, John D. and {Daddi}, Emanuele and {Ho}, Luis C. and {Ding}, Xuheng and {Mercier}, Wilfried and {Faisst}, Andreas L. and {Sheth}, Kartik and {Valentino}, Francesco and {Puglisi}, Annagrazia and {Saito}, Toshiki and {Kakkad}, Darshan and {Ilbert}, Olivier and {Khostovan}, Ali Ahmad and {Liu}, Zhaoxuan and {Tanaka}, Takumi and {Magdis}, Georgios and {Zavala}, Jorge A. and {Tan}, Qinghua and {Kartaltepe}, Jeyhan S. and {Yang}, Lilan and {Koekemoer}, Anton M. and {McKinney}, Jed and {Robertson}, Brant E. and {Jin}, Shuowen and {Hayward}, Christopher C. and {Hirschmann}, Michaela and {Franco}, Maximilien and {Shuntov}, Marko and {Gozaliasl}, Ghassem and {Kaminsky}, Aidan and {Rich}, R. Michael},
        title = "{Clumps as multiscale structures in cosmic noon galaxies}",
      journal = {\mnras},
         year = 2025,
        month = jan,
       volume = {536},
       number = {3},
        pages = {3090-3111},
          doi = {10.1093/mnras/stae2781},
archivePrefix = {arXiv},
       eprint = {2501.03328},
 primaryClass = {astro-ph.GA},
       adsurl = {https://ui.adsabs.harvard.edu/abs/2025MNRAS.536.3090K}
}

@ARTICLE{kalita2024,
       author = {{Kalita}, Boris S. and {Silverman}, John D. and {Daddi}, Emanuele and {Bottrell}, Connor and {Ho}, Luis C. and {Ding}, Xuheng and {Yang}, Lilan},
        title = "{A Rest-frame Near-IR Study of Clumps in Galaxies at 1 < z < 2 Using JWST/NIRCam: Connection to Galaxy Bulges}",
      journal = {\apj},
         year = 2024,
        month = jan,
       volume = {960},
       number = {1},
          eid = {25},
        pages = {25},
          doi = {10.3847/1538-4357/acfee4},
archivePrefix = {arXiv},
       eprint = {2309.05737},
 primaryClass = {astro-ph.GA},
       adsurl = {https://ui.adsabs.harvard.edu/abs/2024ApJ...960...25K}
}

@ARTICLE{2009MNRAS.397L..64A,
       author = {{Agertz}, Oscar and {Teyssier}, Romain and {Moore}, Ben},
        title = "{Disc formation and the origin of clumpy galaxies at high redshift}",
      journal = {\mnras},
         year = 2009,
        month = jul,
       volume = {397},
       number = {1},
        pages = {L64-L68},
          doi = {10.1111/j.1745-3933.2009.00685.x},
archivePrefix = {arXiv},
       eprint = {0901.2536},
 primaryClass = {astro-ph.GA},
       adsurl = {https://ui.adsabs.harvard.edu/abs/2009MNRAS.397L..64A}
}

@ARTICLE{2009ApJ...703..785D,
       author = {{Dekel}, Avishai and {Sari}, Re'em and {Ceverino}, Daniel},
        title = "{Formation of Massive Galaxies at High Redshift: Cold Streams, Clumpy Disks, and Compact Spheroids}",
      journal = {\apj},
         year = 2009,
        month = sep,
       volume = {703},
       number = {1},
        pages = {785-801},
          doi = {10.1088/0004-637X/703/1/785},
archivePrefix = {arXiv},
       eprint = {0901.2458},
 primaryClass = {astro-ph.GA},
       adsurl = {https://ui.adsabs.harvard.edu/abs/2009ApJ...703..785D}
}

@ARTICLE{2010MNRAS.409.1088B,
       author = {{Bournaud}, Fr{\'e}d{\'e}ric and {Elmegreen}, Bruce G. and {Teyssier}, Romain and {Block}, David L. and {Puerari}, Iv{\^a}nio},
        title = "{ISM properties in hydrodynamic galaxy simulations: turbulence cascades, cloud formation, role of gravity and feedback}",
      journal = {\mnras},
         year = 2010,
        month = dec,
       volume = {409},
       number = {3},
        pages = {1088-1099},
          doi = {10.1111/j.1365-2966.2010.17370.x},
archivePrefix = {arXiv},
       eprint = {1007.2566},
 primaryClass = {astro-ph.CO},
       adsurl = {https://ui.adsabs.harvard.edu/abs/2010MNRAS.409.1088B}
}

@ARTICLE{2017MNRAS.464..635M,
       author = {{Mandelker}, Nir and {Dekel}, Avishai and {Ceverino}, Daniel and {DeGraf}, Colin and {Guo}, Yicheng and {Primack}, Joel},
        title = "{Giant clumps in simulated high- z Galaxies: properties, evolution and dependence on feedback}",
      journal = {\mnras},
         year = 2017,
        month = jan,
       volume = {464},
       number = {1},
        pages = {635-665},
          doi = {10.1093/mnras/stw2358},
archivePrefix = {arXiv},
       eprint = {1512.08791},
 primaryClass = {astro-ph.GA},
       adsurl = {https://ui.adsabs.harvard.edu/abs/2017MNRAS.464..635M}
}

@ARTICLE{2019MNRAS.489.2792Z,
       author = {{Zanella}, A. and {Le Floc'h}, E. and {Harrison}, C.~M. and {Daddi}, E. and {Bernhard}, E. and {Gobat}, R. and {Strazzullo}, V. and {Valentino}, F. and {Cibinel}, A. and {S{\'a}nchez Almeida}, J. and {Kohandel}, M. and {Fensch}, J. and {Behrendt}, M. and {Burkert}, A. and {Onodera}, M. and {Bournaud}, F. and {Scholtz}, J.},
        title = "{A contribution of star-forming clumps and accreting satellites to the mass assembly of z {\ensuremath{\sim}} 2 galaxies}",
      journal = {\mnras},
         year = 2019,
        month = oct,
       volume = {489},
       number = {2},
        pages = {2792-2818},
          doi = {10.1093/mnras/stz2099},
archivePrefix = {arXiv},
       eprint = {1907.12136},
 primaryClass = {astro-ph.GA},
       adsurl = {https://ui.adsabs.harvard.edu/abs/2019MNRAS.489.2792Z}
}

@ARTICLE{Claeyssens2025,
       author = {{Claeyssens}, Ad{\'e}la{\"\i}de and {Adamo}, Angela and {Messa}, Matteo and {Dessauges-Zavadsky}, Miroslava and {Richard}, Johan and {Kramarenko}, Ivan and {Matthee}, Jorryt and {Naidu}, Rohan P.},
        title = "{Tracing star formation across cosmic time at tens of parsec-scales in the lensing cluster field Abell 2744}",
      journal = {\mnras},
         year = 2025,
        month = mar,
       volume = {537},
       number = {3},
        pages = {2535-2558},
          doi = {10.1093/mnras/staf058},
archivePrefix = {arXiv},
       eprint = {2410.10974},
 primaryClass = {astro-ph.GA},
       adsurl = {https://ui.adsabs.harvard.edu/abs/2025MNRAS.537.2535C}
}

@article{Windhorst_2022,
   title={JWST PEARLS. Prime Extragalactic Areas for Reionization and Lensing Science: Project Overview and First Results},
   volume={165},
   ISSN={1538-3881},
   url={http://dx.doi.org/10.3847/1538-3881/aca163},
   DOI={10.3847/1538-3881/aca163},
   number={1},
   journal={The Astronomical Journal},
   publisher={American Astronomical Society},
   author={Windhorst, Rogier A. and Cohen, Seth H. and Jansen, Rolf A. and Summers, Jake and Tompkins, Scott and Conselice, Christopher J. and Driver, Simon P. and Yan, Haojing and Coe, Dan and Frye, Brenda and Grogin, Norman and Koekemoer, Anton and Marshall, Madeline A. and O’Brien, Rosalia and Pirzkal, Nor and Robotham, Aaron and Ryan, Russell E. and Willmer, Christopher N. A. and Carleton, Timothy and Diego, Jose M. and Keel, William C. and Porto, Paolo and Redshaw, Caleb and Scheller, Sydney and Wilkins, Stephen M. and Willner, S. P. and Zitrin, Adi and Adams, Nathan J. and Austin, Duncan and Arendt, Richard G. and Beacom, John F. and Bhatawdekar, Rachana A. and Bradley, Larry D. and Broadhurst, Tom and Cheng, Cheng and Civano, Francesca and Dai, Liang and Dole, Hervé and D’Silva, Jordan C. J. and Duncan, Kenneth J. and Fazio, Giovanni G. and Ferrami, Giovanni and Ferreira, Leonardo and Finkelstein, Steven L. and Furtak, Lukas J. and Gim, Hansung B. and Griffiths, Alex and Hammel, Heidi B. and Harrington, Kevin C. and Hathi, Nimish P. and Holwerda, Benne W. and Honor, Rachel and Huang, Jia-Sheng and Hyun, Minhee and Im, Myungshin and Joshi, Bhavin A. and Kamieneski, Patrick S. and Kelly, Patrick and Larson, Rebecca L. and Li, Juno and Lim, Jeremy and Ma, Zhiyuan and Maksym, Peter and Manzoni, Giorgio and Meena, Ashish Kumar and Milam, Stefanie N. and Nonino, Mario and Pascale, Massimo and Petric, Andreea and Pierel, Justin D. R. and Carmen Polletta, Maria del and Röttgering, Huub J. A. and Rutkowski, Michael J. and Smail, Ian and Straughn, Amber N. and Strolger, Louis-Gregory and Swirbul, Andi and Trussler, James A. A. and Wang, Lifan and Welch, Brian and B. Wyithe, J. Stuart and Yun, Min and Zackrisson, Erik and Zhang, Jiashuo and Zhao, Xiurui},
   year={2022},
   month=dec, pages={13} }

@article{10.1093/mnras/stv300,
    author = {Walker, D. L. and Longmore, S. N. and Bastian, N. and Kruijssen, J. M. D. and Rathborne, J. M. and Jackson, J. M. and Foster, J. B. and Contreras, Y.},
    title = {Tracing the conversion of gas into stars in Young Massive Cluster Progenitors},
    journal = {Monthly Notices of the Royal Astronomical Society},
    volume = {449},
    number = {1},
    pages = {715-725},
    year = {2015},
    month = {03},
    issn = {0035-8711},
    doi = {10.1093/mnras/stv300},
    url = {https://doi.org/10.1093/mnras/stv300},
    eprint = {https://academic.oup.com/mnras/article-pdf/449/1/715/4141179/stv300.pdf},
}

@ARTICLE{1996A&AS..117..393B,
       author = {{Bertin}, E. and {Arnouts}, S.},
        title = {\textcolor{red}{SExtractor: Software for source extraction.}},
      journal = {\aaps},
         year = 1996,
        month = jun,
       volume = {117},
        pages = {393-404},
          doi = {10.1051/aas:1996164},
       adsurl = {https://ui.adsabs.harvard.edu/abs/1996A&AS..117..393B}
}

\clearpage
\onecolumn
\begin{appendix}
\section{Alternative Morphologies}
\label{app:shape}

\begin{figure}[h]
      \centering		
      \includegraphics[width=\linewidth]{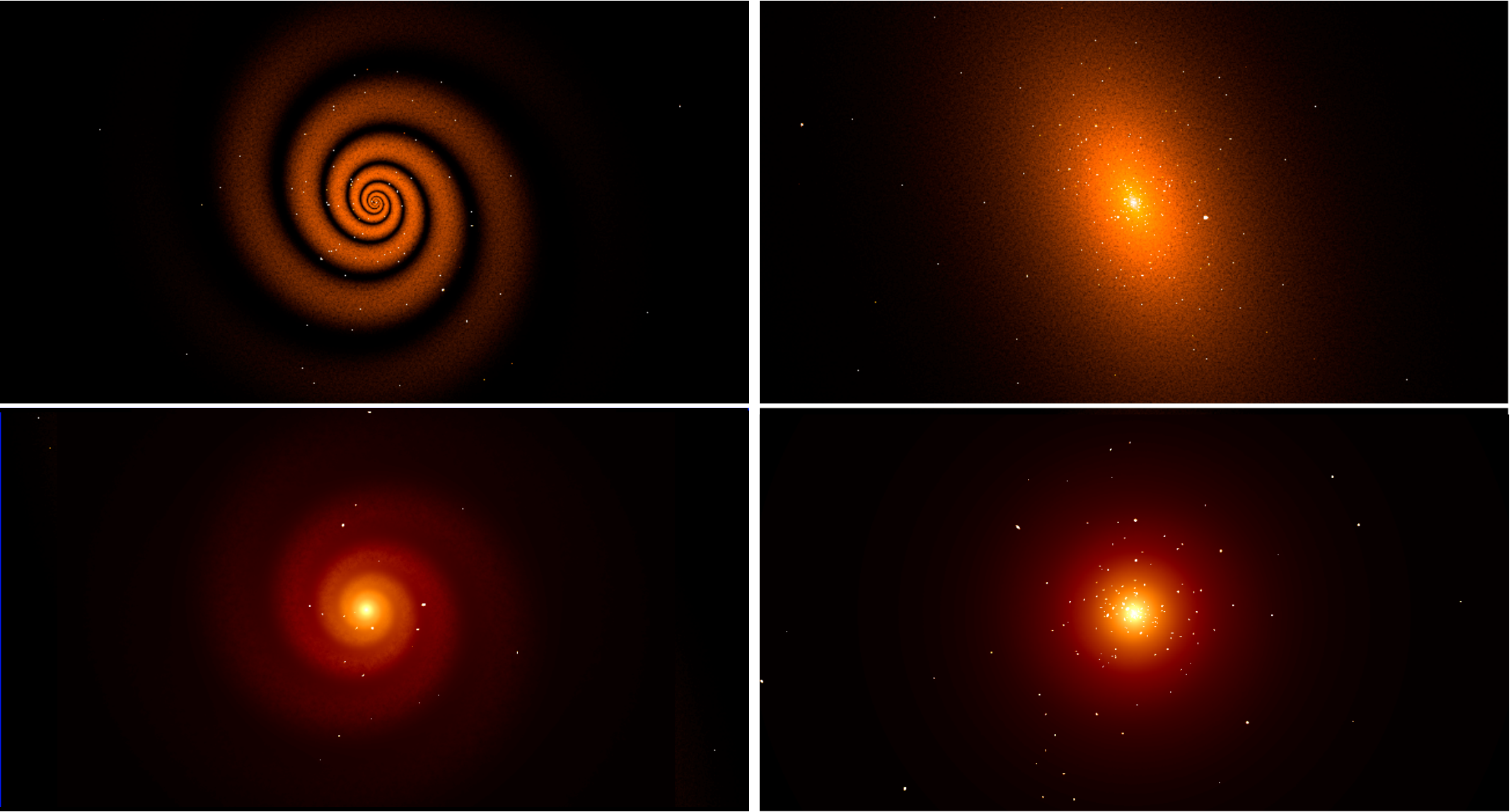}
       \caption{Four examples of alternative morphologies for the source galaxy. Top left: a spiral galaxy composed exclusively of spiral arms, with a clump fraction $Q = 80\%$; top right: an elliptical-like galaxy with a single ellipsoidal component, and $Q = 80\%$; bottom left: a spiral galaxy including both spiral arms and a central spheroidal bulge, with $Q = 1\%$; bottom right: a spiral galaxy composed of an exponential disk and a central spheroidal bulge, with $Q = 90\%$. All models are shown at an angular resolution of 0.25 mas/pixel}
	\label{shape_app}
\end{figure}

\clearpage
\section{Keywords}
\label{app:keywords}
\begin{table}[h!]
    \centering
    \captionsetup{position=above}
    \caption{List of the parameters used in the pipeline.}
    \label{parameters}

    \begin{minipage}[t]{0.49\textwidth}
    \vfill
    \resizebox{\textwidth}{!}
    {\fontsize{9pt}{10pt}\selectfont
    \begin{tabularx}{\textwidth}{||X | >{\raggedright\arraybackslash}p{6cm}||}
     \hline
        \multicolumn{2}{|c|}{Source: galaxy} \\         
        \hline
         $Mass\_bul$ & Stellar mass of the bulge [$M_{\odot}$]\\
         \hline
         $Re\_bul$ & Effective radius of the bulge [arcsec]\\
         \hline
         $n\_bul$ & S\'ersic index of the bulge\\
         \hline
         $q\_bul$ & Axis ratio of the bulge\\
         \hline
         $pa\_bul$ & Position angle of the bulge\\
         \hline
         $Mass\_disk$ & Stellar mass of the disk [$M_{\odot}$]\\
         \hline
         $Re\_disk$ & Effective radius of the disk [arcsec]\\
         \hline
         $n\_disk$ & S\'ersic index of the disk\\
         \hline
         $q\_disk$ & Axis ratio of the disk\\
         \hline
         $pa\_disk$ & Position angle of the disk\\
         \hline
         $N_a$ & Number of spiral arms\\
         \hline
         $\phi_d$ & Phase angle of the spiral arms\\
         \hline
         $\alpha$ & Parameter describing how the arms are wrapped around the bulge\\
         \hline         
        $SFH\_host$ & Star Formation History of the host galaxy (input parameter for the Yggdrasil Models)\\
        \hline
        $age\_disk$ & Age of the disk (input parameter for Yggdrasil)\\
        \hline
        $age\_bul$ & Age of the bulge (input parameter for Yggdrasil)\\
        \hline
         $Z$ &  Metallicity (input parameter for Yggdrasil)\\
        \hline
        $IMF$ & Initial Mass Function (input parameter for Yggdrasil)\\
        \hline
         $z\_s$ & Source redshift \\
         \hline
        \hline
         \multicolumn{2}{|c|}{Source: star clusters} \\
         \hline         
         $fsub$ & Fraction of the host galaxy stellar mass in clumps\\
         \hline
         $\beta$ & Power-law slope of the cluster mass function (CMF)\\
         \hline
         $trunc\_mass$ & High-mass truncation of the CMF [$M_{\odot}$]\\
         \hline
         $\delta$ & Shape of the exponential cut-off in the CMF\\
         \hline
         $\gamma$ & Steepness of the exponential cut-off in the CMF\\
         \hline
         $mass\_min$, $mass\_max$  & Lower and upper bounds for the CMF sampling [$M_{\odot}$]\\
         \hline
         $\alpha\_t$ & Logarithmic slope of the cluster age function  \\
         \hline
         $t\_min$,$t\_max$ & Lower and upper bounds for sampling the ages of the clumps\\
         \hline
        $SFH\_cl$ & Star Formation History of the clumps(input parameter for Yggdrasil)\\
         \hline
    \end{tabularx}
    }
    \vfill
    \end{minipage}
    \hspace{-0.05cm}
    \begin{minipage}[t]{0.49\textwidth}
    \vfill
    \resizebox{\textwidth}{!}
    {\fontsize{9pt}{10pt}\selectfont
    \begin{tabularx}{\textwidth}{||X | >{\raggedright\arraybackslash}p{4.7cm}||}
        \hline
         \multicolumn{2}{|c|}{Image configuration} \\
         \hline         $pix\_scale\_unl$ & The pixel scale of the unlensed images [arcsec/pix]\\
         \hline
         $pix\_scale$ & The pixel scale of the lensed images [arcsec/pix]\\
         \hline
         $sizex$,$sizey$ & Two-element tuples indicating the width of the image along the x and y-axis on the lens plane, given in arcsec from the lens center\\
         \hline
         $size\_unl$ & Image dimension on the source plane [arcsec]\\
        \hline\hline
         \multicolumn{2}{|c|}{Lens} \\
         \hline
         $zl$ & Cluster redshift\\
         \hline
         $coords$ & String with the cluster coordinates in degrees\\
         \hline
         $gal\_x$, $gal\_y$ & Source position along the x and y-axis on the source plane, given in arcsec from the lens center\\
        \hline\hline
        \multicolumn{2}{|c|}{Filters and noise}\\
        \hline
        $\lambda_{red}$, $\lambda_{green}$, $\lambda_{blue}$ & Effective observed wavelengths used to create an RGB high resolution image [\AA]\\
        \hline
        $ZP\_instr$ & ZP for the selected filter [ABmag]\\
        \hline
        $Texp\_instr$ & Exposure time for the selected filter [s]\\
        \hline
        $Flux\_sky\_instr$ & Sky flux for the selected filter [counts/s/pix]\\
        \hline
        $pix\_scl\_instr$ & Native pixel size of the selected filter [arcsec/pix] \\
        \hline
        $pix\_scl\_PSF\_instr$ & Pixel size of the PSF for the selected filter [arcsec/pix]\\
        \hline
    \end{tabularx}
    }
    \vfill
    
    \end{minipage}
\end{table}

\end{appendix}
\end{document}